\documentclass{llncs}

\usepackage{epsfig}
\usepackage{subcaption}
\usepackage{calc}
\usepackage{amssymb}
\usepackage{amstext}
\usepackage{amsmath}
\usepackage{algorithm}
\usepackage{algpseudocode}
\usepackage{multicol}
\usepackage{pslatex}
\usepackage[bottom]{footmisc}
\usepackage{booktabs}
\usepackage{tabularx}
\usepackage{threeparttable}
\usepackage{hyperref}
\usepackage{cleveref}
\usepackage{balance}
\usepackage{float}
\usepackage{placeins}
\usepackage{xcolor}

\graphicspath{{./images/}}

\newif\ifanonymous
\anonymousfalse

\newcommand{\anonymize}[1]{%
  \ifanonymous
    Anonymized for review%
  \else
    #1%
  \fi
}

\begin{document}

\title{Beyond Symmetry: Asymmetric Matching with Transformers for Collective Entity Resolution}

\author{\anonymize{Andrei Olar\inst{1}\orcidID{0009-{}0006-{}7913-{}9276}}}

\institute{
    Faculty of Mathematics and Computer Science, Babe\c{s}-Bolyai University
    \email{andrei.olar@ubbcluj.ro}
}
\maketitle

\begin{abstract}
    A core assumption of entity resolution (ER) is its symmetric nature, which rests on
    a principle so self-evident it borders on tautology: if a refers to the same object
    as b, then b must also refer to the same object as a. However, this global property
    of symmetry does not necessarily extend to each of the complex decision processes
    within ER systems. In cases involving vague or ambiguous information, enforcing
    symmetry can be detrimental to performance, an issue exemplified by the ``bad
    triplet problem''. This paper investigates the effects of input-order asymmetry
    within a collective ER framework that integrates distinct matching and clustering
    stages. The input order-aware matching stage utilizes pre-trained transformers to
    explicitly model and exploit asymmetry, making it possible to evaluate its effect on
    clustering performance across multiple benchmark datasets. Our results demonstrate
    that asymmetry is not only measurable but also introduces a significant skew in the
    final entity clusters, highlighting the need to re-evaluate supervised training
    methodologies for matchers, particularly the development of direction-aware
    training paradigms that can preserve, rather than neutralize, the inherent asymmetry
    in matching tasks.

\end{abstract}

\section{Introduction}\label{sec:e2e-transformers-introduction}

The concept of coreference is inherently symmetric: if \emph{a} refers to the same
thing as \emph{b}, then \emph{b} refers to the same thing as \emph{a}. In Entity
Resolution (ER), this logical truism has traditionally dictated the design of system
architectures. While coreference is symmetric after the fact, some of the steps leading
to the coreference decision need not be. This is particularly true for modern ER
matchers built on sequence-aware transformer architectures, where the order of input
records can influence the resulting similarity score. For instance, a model comparing a
sparse record like ``John Smith'' to a rich record ``J. Smith, MIT, PhD'' may produce a
different confidence score than when the inputs are reversed, as the contextual cues
for the matching decision differ in each direction.

The implicit assumption of symmetry can mask valuable information and, in cases
involving ambiguous data, even degrade performance—a phenomenon related to the ``bad
triplet problem''~\cite{badtriplet2008} where enforcing transitivity leads to incorrect
entity clusters. This paper challenges the universality assumption of symmetry
throughout the ER process by investigating the effects of input-order asymmetry within
a collective ER framework~\cite{bhatget2007}. We build upon preliminary
findings~\cite{olar2025} that first identified and measured this phenomenon in a
RoBERTa-based~\cite{roberta2019} matcher using the Ditto~\cite{ditto2020} architecture.
Here, we significantly extend that work by demonstrating that asymmetry is not an
artifact of a single implementation but a general characteristic of transformer-based
matchers. We expand the experimental scope to include DistilBERT~\cite{distilbert2019},
TinyBERT~\cite{tinybert2020}, and DeBERTa~\cite{deberta2023}, revealing consistent
asymmetric behaviors across these diverse models.

The core of our investigation lies in a two-stage ER pipeline comprising an order-aware
matching stage and a subsequent clustering stage. By constructing a directed reference
graph~\cite{bhatget2006} from the matcher's output, we explicitly model and exploit the
asymmetry in pairwise comparisons. We then analyze how this directed information is
processed by various clustering algorithms, including weakly connected
components~\cite{wcc1972}, CENTER~\cite{center2000}, Markov
Clustering~\cite{markovclusters2008}, and a PageRank-based approach
(ACL)~\cite{gacl2024}. Our results show that asymmetry is not only measurable but also
introduces a significant and predictable skew in the final entity clusters. This work
highlights the need to re-evaluate how to model ER matcher output in the context of
using modern matchers within a graphical collective ER framework, particularly
regarding the trade-offs between different clustering strategies.

To this end, the remainder of this paper is organized as follows. We first survey
related work before detailing our experimental methodology. We then present our
expanded results, analyzing the prevalence of asymmetry across multiple transformer
models and its downstream effect on clustering performance. We conclude with a
discussion of our findings and their implications for future work in collective entity
resolution.

\section{Related Work}\label{sec:sec:e2e-transformers-related-work}

Our work is positioned at the intersection of three areas: the foundational principles
of entity resolution, the architectural evolution of deep learning-based matchers, and
the graph-based collective ER model. We structure our review to highlight a critical,
yet overlooked, architectural side-effect of modern ER models: input-order asymmetry.

\subsection{The Implicit Assumption of Symmetry in Traditional and Collective ER}

The concept of symmetry is deeply embedded in the theory and practice of ER.
Traditional pairwise similarity functions, such as Jaccard or Levenshtein distance, are
inherently symmetric by definition. Early probabilistic models of ER~\cite{fs1969} and
their subsequent refinements preserve this symmetry, typically by composing similarity
functions to determine a final score where $sim(a, b)$ equals $sim(b,
a)$~(see~\cite{winkler2014} for a detailed review). This property is not merely a
convenience but a reflection of the symmetric nature of object identity
itself~\cite{christen2012data}. When machine learning was first applied to ER, it was
often for the task of learning symmetric similarity functions~\cite{bilenko2003}. This
property was preserved by design, for instance, through data augmentation that trains a
model on both $(a, b)$ and $(b, a)$ with the same label, or via architectures like
Siamese networks, which learn a symmetric distance metric by processing inputs through
identical, weight-sharing
subnetworks~\cite{loster2021knowledge,nseen2019,krivosheev2021}.

This foundational assumption of symmetry directly shapes the practice of collective
entity resolution (CER)~\cite{bhatget2007}. CER moves beyond pairwise decisions to
resolve entities based on the global structure of their relationships, a process almost
universally modeled using an \emph{undirected similarity
graph}~\cite{hassanzadehclustering2009,saeediclustering2017}. In this paradigm, records
are represented as nodes, and an edge exists between two nodes if their similarity
score exceeds a certain threshold. The edge is, by definition, undirected because the
similarity score is assumed to be symmetric. Foundational work on reference
graphs~\cite{bhatget2004} and the vast body of literature on ER clustering algorithms
operate on this premise, developing techniques for partitioning undirected graphs into
entity clusters~\cite{bipartitegraphclusters2023,relater2023}.

\subsection{The Evolution to Sequential, Asymmetry-Prone Matchers}

The advent of deep learning introduced more powerful, but also more complex, ER
matchers. Early models like DeepMatcher~\cite{deepmatcher2018} and
DeepER~\cite{deeper2018}, which were often based on Recurrent Neural Networks (RNNs),
were cognizant of their sequential nature. They took explicit steps to mitigate
potential order-effects and enforce symmetry, for example by using attention mechanisms
over attribute summaries (DeepMatcher) or employing Siamese-style architectures to
compare latent representations in a symmetric manner (DeepER).

The subsequent shift to transformer-based models marked a significant architectural
change. Models based on the Ditto architecture~\cite{ditto2020}, in particular, depart
from the Siamese paradigm. Instead of processing each record independently, Ditto
serializes the two records into a single, long sequence, typically formatted as
\texttt{[CLS] attr val [SEP] attr val [SEP]}. This sequence is then fed into a
transformer encoder like BERT to produce a match/non-match classification. While highly
effective, this design is fundamentally sequential and thus inherently sensitive to the
order in which the two records appear in the input sequence. The latest trend of using
Large Language Models (LLMs) for ER often adopts a similar pairwise prompting or
fine-tuning approach~\cite{incontext2025,matchcompareselect2025}. The focus of this
research has been almost exclusively on improving matching accuracy, with systems like
CrossER that use cross-attention to explicitly handle input order being a notable
exception~\cite{crosser2026}.

\subsection{A Gap in the Literature: Underexplored Asymmetry and Directed Clustering}

The architectural evolution from symmetric-by-design models to sequential
transformer-based models reveals a critical gap in the literature. Early deep learning
models \emph{actively worked to enforce symmetry}, whereas many state-of-the-art
transformer and LLM-based models use architectures that are \emph{inherently
asymmetric}. This property and its consequences for the ER process remain largely
unexamined. While recent systems have begun to explore more complex graph structures in
CER, they do not address the issue of transformer-induced asymmetry. For example,
GraphER~\cite{grapher2025} uses GNNs with graph differential dependencies, and
RELATER~\cite{relater2023} uses an unsupervised approach on undirected graphs to remove
incorrect links; neither leverages transformer-based matchers as the source of their
graph's structure.

This underexamined scenario leads to a practical challenge: the vast majority of ER
clustering research is built for undirected graphs. Consequently, practitioners using
modern matchers are faced with a choice: either force the resulting directed graph to
be symmetric (e.g., by averaging edge weights), thereby discarding potentially valuable
directional information, or find themselves unable to use most standard clustering
algorithms.

This paper directly addresses this gap. To our knowledge, it is among the first to (1)
systematically quantify and compare the asymmetry produced by a range of modern
transformer-based matchers, and (2) empirically evaluate the performance of various
clustering algorithms capable of operating on directed graphs when this asymmetric
information is preserved rather than discarded.

\section{Methodology}\label{sec:methodology}

\begin{figure}[h!]
    \centering
    \includegraphics[width=0.9\textwidth]{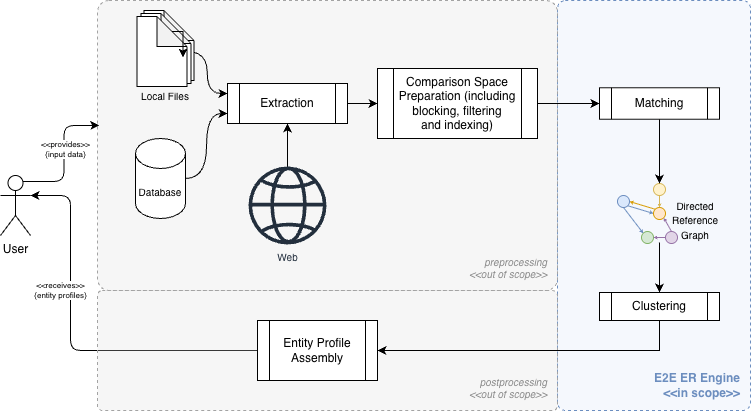}
    \caption{
        The two-stage ER pipeline. The focus of this paper is rendered with a blue
        background. The asymmetric matching stage uses a transformer-based architecture
        to produce a directed reference graph, which is then partitioned by a clustering
        algorithm.
    }\label{fig:e2e-transformers-pipeline}
\end{figure}

To investigate the effects of matcher asymmetry on collective ER, we adopt a two-stage
pipeline. This framework, illustrated in Figure~\ref{fig:e2e-transformers-pipeline},
consists of an \emph{asymmetric matching stage} followed by a \emph{clustering stage}.
The central insight of this approach is to model the matcher's output not as a
symmetric similarity matrix, but as a \textbf{directed, weighted graph} $G=(V, E, w)$.
In this graph, vertices $V$ represent records, and a directed edge $e = (v_i, v_j) \in
E$ exists if the matcher's confidence score $w_{ij} = \text{score}(v_i, v_j)$ exceeds a
threshold $\tau$. The potential for asymmetry means that $w_{ij}$ is not guaranteed to
be equal to $w_{ji}$.

Building on a preliminary study that established this model with a single RoBERTa-based
matcher~\cite{olar2025}, this paper expands the scope of both stages to test the
generality of the asymmetry phenomenon and its downstream consequences.

\begin{figure}[h!]
    \centering
    \includegraphics[width=0.9\textwidth]{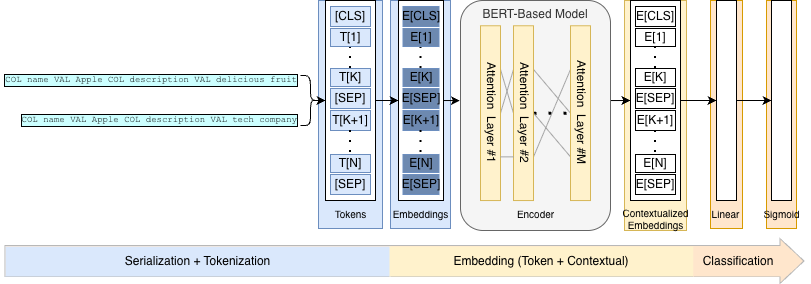}
    \caption{The Ditto architecture. Adapted from Olar (2025)~\cite{olar2025}.}
    \label{fig:e2e-transformers-matcher-architecture}
\end{figure}

\subsection{Asymmetric Matching Stage}\label{subsec:matching_stage}

The matching stage employs transformer models fine-tuned on ER tasks using the Ditto
architecture~\cite{ditto2020} shown in
Figure~\ref{fig:e2e-transformers-matcher-architecture}, which serializes record pairs
into a single sequence for classification. This sequential processing is the primary
source of the input-order sensitivity we investigate. To demonstrate that asymmetry is
a general property of this approach, we expand our experiments beyond the original
RoBERTa model~\cite{roberta2019} to include the additional transformers with diverse
characteristics. The models and their key architectural parameters are detailed in
Table~\ref{tbl:e2e-transformers-models}. These models were chosen to represent a
spectrum of sizes and architectural variations, from the large RoBERTa and DeBERTa
models to the more compact DistilBERT and TinyBERT.

\begin{table}[h!]
    \centering
    \begin{tabular}{lcccc}
        \hline
        \textbf{Model}                                & \textbf{Hidden Dim.} & \textbf{Layers} & \textbf{Heads} & \textbf{Total Params.} \\ \hline
        RoBERTa-base~\cite{roberta2019}               & 768                  & 12              & 12             & 125M                   \\
        DeBERTa-v3-base~\cite{deberta2023}            & 768                  & 12              & 12             & 184M                   \\
        DistilBERT-base-uncased~\cite{distilbert2019} & 768                  & 6               & 12             & 66M                    \\
        TinyBERT-6L-768d~\cite{tinybert2020}          & 768                  & 6               & 12             & 66M                    \\ \hline
    \end{tabular}
    \caption{Key architectural parameters of the transformer models used in the matching stage.}\label{tbl:e2e-transformers-models}
\end{table}

By evaluating across this spectrum of models, we can assess whether asymmetry is a
consistent feature rather than an artifact of a single model's size or architecture.
The outcome from running the matcher is a directed graph. Depending on the parameter
input order, the matcher will output either a \emph{normal} reference graph or a
\emph{reversed} reference graph~\cite{olar2025}.

\subsection{Clustering Stage}\label{subsec:clustering_stage}

The clustering stage takes the directed graph $G$ from the matcher and partitions its
vertices into clusters representing real-world entities. We evaluate four algorithms
chosen to represent different strategies for handling (or ignoring) edge
directionality. Following the principles for robust evaluation outlined by Hassanzadeh
et al.~\cite{hassanzadehclustering2009}, our selection prioritizes
\textbf{unconstrained algorithms}—those that do not require a predefined number of
clusters ($k$) or extensive, dataset-specific parameter tuning. This focus on
simplicity and reproducibility allows for a direct assessment of how each algorithm's
core strategy interacts with the asymmetric graph structure produced by the matcher.

\paragraph{Direction-Agnostic Baseline.}
\textbf{Weakly Connected Components (WCC)}~\cite{wcc1972} serves as our baseline. It
forms clusters by grouping all nodes connected by a path, regardless of edge direction,
and is a widely adopted baseline in ER literature for evaluating performance on the
underlying undirected graph~\cite{saeediclustering2017,bipartitegraphclusters2023}. By
discarding all directional information from the matcher, WCC allows us to measure the
impact of this information loss. Our experiments use the canonical implementation
provided by the \texttt{networkx} Python package~\cite{networkx}.

\paragraph{Direction-Aware Heuristics.}
\textbf{CENTER Clustering}~\cite{center2000} is a deterministic heuristic designed for
large-scale ER. It forms star-shaped clusters by designating nodes as centers and
grouping all nodes reachable via an outgoing edge. While related heuristics have been
applied to undirected similarity graphs~\cite{hassanzadehclustering2009}, CENTER's
design is particularly suited for directed graphs, making it highly sensitive to the
matcher's asymmetry. Despite its conceptual simplicity, a canonical implementation for
directed graphs is not readily available. We therefore implement our own variant,
detailed in Algorithm~\ref{alg:pc}, which follows an iterative \emph{Parent CENTER} (PC)
approach. In this variant, each node is repeatedly assigned to the root of the cluster
belonging to its most similar predecessor until the assignments stabilize.

\begin{algorithm}
    \caption{Parent CENTER (PC) Clustering}\label{alg:pc}
    \begin{algorithmic}[1]
        \State \textbf{Input:} Directed graph $G=(V, E, w)$
        \State \textbf{Output:} A partition $C$ of the vertices $V$

        \Function{ParentCenter}{$G$}
        \State $parents \gets \text{new map where } parents[v] \gets v \text{ for all } v \in V$
        \State $updated \gets \text{true}$
        \While{$updated$}
        \State $updated \gets \text{false}$
        \State $new\_parents \gets \text{copy}(parents)$
        \For{each node $v \in V$}
        \State $best\_pred \gets v$
        \State $max\_w \gets -1$
        \For{each predecessor $u$ of $v$}
        \If{$w(u, v) > max\_w$}
        \State $max\_w \gets w(u, v)$
        \State $best\_pred \gets u$
        \EndIf
        \EndFor
        \If{$best\_pred \neq v$}
        \State $root \gets \text{FindRoot}(parents, best\_pred)$
        \If{$new\_parents[v] \neq root$}
        \State $new\_parents[v] \gets root$
        \State $updated \gets \text{true}$
        \EndIf
        \EndIf
        \EndFor
        \State $parents \gets new\_parents$
        \EndWhile
        \State \Return group nodes in $V$ by their final parent in $parents$
        \EndFunction
        \Statex
        \Function{FindRoot}{$parents, v$}
        \While{$parents[v] \neq v$}
        \State $parents[v] \gets parents[parents[v]]$ \Comment{Path compression}
        \State $v \gets parents[v]$
        \EndWhile
        \State \Return $v$
        \EndFunction
    \end{algorithmic}
\end{algorithm}

\paragraph{Flow-Based Clustering.}
The first of two methods that leverage global graph properties is \textbf{Markov
Clustering (MCL)}~\cite{markovclusters2008}, an algorithm that simulates random walks
(i.e., flow) on the graph. By iteratively applying \emph{expansion} (modeling flow
spread) and \emph{inflation} (strengthening strong currents), MCL naturally converges
on dense regions that form the final clusters. Because the underlying random walk
simulation respects edge direction, the algorithm is inherently capable of processing
the asymmetric signal from the matcher. For our experiments, we use the canonical
implementation provided by the \texttt{markov-clustering} Python
package~\cite{markovclusters2025}.

\paragraph{Modularity-Based Clustering.}
To address the inherent asymmetry in the entity resolution graph, we employ the
\textbf{Leiden (LEI)} algorithm~\cite{leiden2019}, optimized for the directed
modularity function proposed by Leicht and Newman~\cite{leicht2008community}. Unlike
standard modularity, which assumes undirected interactions, this formulation
generalizes the null model to account for the probability of a directed edge existing
from node $i$ to node $j$ based on their respective out-degrees and in-degrees. The
directed modularity $Q$ is defined as:

\begin{equation}
    Q = \frac{1}{m} \sum_{i,j} \left( A_{ij} - \frac{k_i^{out}k_j^{in}}{m} \right) \delta(c_i, c_j)
\end{equation}

\noindent where $m$ is the total number of directed edges in the network, $A_{ij}$
represents the adjacency matrix (where $A_{ij}=1$ if there is a link from $i$ to $j$,
and 0 otherwise), $k_i^{out}$ is the out-degree of node $i$, and $k_j^{in}$ is the
in-degree of node $j$. The function $\delta(c_i, c_j)$ is the Kronecker delta, which
equals 1 if nodes $i$ and $j$ belong to the same community (cluster) and 0 otherwise.
By optimizing this specific objective function, the clustering process respects the
directional flow of identity information modeled by the transformer matchers.
In our experiments, we utilize the high-performance implementation provided by the
\texttt{leidenalg}~\cite{leidenalg2025} library, built upon the core \texttt{igraph}
framework~\cite{igraph2006}.

\begin{algorithm}[t]
    \caption{Directed Spectral Clustering with Adaptive $k$}\label{alg:spectral}
    \begin{algorithmic}[1]
        \Require Adjacency matrix $A \in \mathbb{R}^{N \times N}$, Teleportation $\alpha$, Threshold $\epsilon$
        \Ensure Cluster assignments $\mathcal{C}$ for all nodes

        \State $\mathcal{C} \leftarrow \text{Array}(N)$
        \State Components $\leftarrow$ \Call{WCC}{$A$}

        \For{each subgraph $G_s$ in Components}
        \State $A_s \leftarrow$ Adjacency of $G_s$
        \State $D_{out} \leftarrow \text{diag}(\sum_j A_{s,ij})$
        \State $P \leftarrow D_{out}^{-1} A_s$ \Comment{Transition Matrix}

        \State \textbf{Step 1: Stationary Distribution (Power Iteration)}
        \State $\pi \leftarrow \text{InitializeUniform}(|G_s|)$
        \Repeat
        \State $\pi_{new} \leftarrow \alpha P^T \pi + (1-\alpha)/|G_s|$
        \Until{$||\pi_{new} - \pi|| < \epsilon$}
        \State $\Pi \leftarrow \text{diag}(\pi)$

        \State \textbf{Step 2: Directed Laplacian Construction~\cite{chung2005laplacians}}
        \State $S \leftarrow \Pi^{1/2} P \Pi^{-1/2}$
        \State $\mathcal{L} \leftarrow I - \frac{1}{2}(S + S^T)$

        \State \textbf{Step 3: Eigendecomposition}
        \State $\lambda_1 \dots \lambda_m, u_1 \dots u_m \leftarrow \text{SmallestEigs}(\mathcal{L})$

        \State \textbf{Step 4: Eigengap Heuristic~\cite{vonluxburg2007tutorial}}
        \State $\delta_i \leftarrow |\lambda_{i+1} - \lambda_i|$
        \State $k \leftarrow \arg\max_{i \ge 1}(\delta_i) + 1$ \Comment{Find stable spectral gap}

        \State \textbf{Step 5: Clustering}
        \State $U \leftarrow [u_1, \dots, u_k]$
        \State $U_{norm} \leftarrow$ Normalize rows of $U$ to unit length
        \State $C_s \leftarrow \text{KMeans}(U_{norm}, k)$
        \State Update $\mathcal{C}$ with $C_s$
        \EndFor
        \State \Return $\mathcal{C}$
    \end{algorithmic}
\end{algorithm}

\paragraph{Directed Spectral Clustering.}
While the Leiden algorithm optimizes local modularity, \textbf{spectral clustering
(SC)} captures global structural properties through the eigenvalues of the graph
Laplacian. However, traditional spectral formulations are primarily designed for
undirected graphs, relying on the symmetry of the Laplacian to guarantee real-valued
eigenvalues and orthogonal eigenvectors~\cite{ng2002spectral,malliaros2013clustering}.
Applying these methods directly to asymmetric matrices often results in complex spectra
that are difficult to interpret for clustering. To leverage global spectral embeddings
while respecting the inherent directionality of our matching data, we implement a
directed spectral clustering pipeline rooted in the random walk
Laplacian~\cite{chung2005laplacians}.

Algorithm~\ref{alg:spectral} first decomposes the graph into weakly connected
components (WCC) to ensure scalability. Within each component, we model the ``flow'' of
identity using a random walk transition matrix $P = D^{-1}A$. We compute the stationary
distribution $\Pi$ of this random walk via power iteration with teleportation,
analogous to PageRank~\cite{pagerank1999}. This distribution allows us to construct a
directed Laplacian matrix $\mathcal{L}$ that is symmetric and positive semi-definite,
yet fully derived from the asymmetric transition probabilities. By solving for the $k$
smallest eigenvectors of $\mathcal{L}$, we generate a low-dimensional embedding that
preserves the directional influence of the entities. The number of clusters $k$ is
determined dynamically via the eigengap heuristic~\cite{vonluxburg2007tutorial}, and
the final partition is extracted by applying $k$-means to the normalized spectral
embedding.

\subsection{Evaluation Protocol}\label{subsec:evaluation}

Our evaluation protocol is designed to measure asymmetry at the matcher level and its
subsequent impact on clustering performance.

\paragraph{Matcher Evaluation.}
We quantify asymmetry using a skew metric defined in our prior work~\cite{olar2025},
which measures the directional imbalance in similarity scores. We also evaluate the
overall matching performance for both input directions using the standard $F_1$ score
to ensure our models are comparable to the original Ditto baselines.

\paragraph{Clustering Evaluation.}
We first establish a ground truth partition by converting the pairwise ground truth
labels into a set of equivalence classes using a union-find data
structure~\cite{tarjanunionfind1979,olar2025}. Each clustering algorithm is run on both
the \emph{normal} and the \emph{reversed} directed graphs produced by the
matcher~\cite{olar2025}. We then compare the resulting partitions against the ground
truth using a suite of metrics: the Adjusted Rand Index~\cite{adjrand2001}, the
Talburt-Wang Index~\cite{tal2007}, the Pairwise Comparison
Measure~\cite{menestrinaermetrics2010}, and the Cluster Comparison
Measure~\cite{huang2006efficient}.

\section{Experiments}\label{sec:experiments}

The experiments are designed to empirically validate the methodology outlined in
Section~\ref{sec:methodology}. We aim to (1) measure and compare the input-order
asymmetry across four different transformer architectures, and (2) evaluate the
downstream impact of this asymmetry on several distinct graph clustering algorithms.
The experimental design, including datasets and core training procedures, follows our
preliminary work~\cite{olar2025}.

\subsection{Datasets and Metrics}

\paragraph{Datasets.}
We use eight publicly available benchmark datasets, previously used in the evaluation
of Ditto~\cite{ditto2020} and DeepMatcher~\cite{deepmatcher2018}. As detailed
previously~\cite{olar2025}, these are all Clean-Clean ER datasets~\cite{e2e2020}, where
duplicates exist only between the two sources. The characteristics of these datasets
are summarized in Table~\ref{tbl:datasets}. Each dataset provides standard training,
validation, and test splits, which are used without modification.

\begin{table}[h!]
    \centering
    \caption{
        Benchmark dataset characteristics. $|\mathcal{CS}|$ is the size of the
        comparison space and $|\mathcal{T}|$ is the number of true matches.
    }\label{tbl:datasets}
    \resizebox{\columnwidth}{!}{
        \begin{tabular}{llccccrrr}
            \hline
            \textbf{Name}  & \textbf{Domain} & \textbf{$|L|$} & \textbf{$|A_L|$} & \textbf{$|R|$} & \textbf{$|A_R|$} & \textbf{$|L \times R|$} & \textbf{$|\mathcal{CS}|$} & \textbf{$|\mathcal{T}|$} \\
            \hline
            Abt-Buy        & e-commerce      & 1081           & 4                & 1092           & 4                & 1180452                 & 9575                      & 1028                     \\
            Amazon-Google  & software        & 1363           & 4                & 3226           & 4                & 4397038                 & 11460                     & 1167                     \\
            Beer           & beer            & 4345           & 5                & 3000           & 5                & 13035000                & 450                       & 68                       \\
            DBLP-ACM       & academic        & 2616           & 5                & 2294           & 5                & 6001104                 & 12363                     & 2220                     \\
            DBLP-Scholar   & academic        & 2616           & 5                & 64263          & 5                & 168112008               & 28707                     & 5347                     \\
            Fodors-Zagat   & hospitality     & 533            & 7                & 331            & 7                & 176423                  & 946                       & 110                      \\
            iTunes-Amazon  & music           & 6907           & 9                & 55923          & 9                & 386260161               & 539                       & 132                      \\
            Walmart-Amazon & electronics     & 2554           & 6                & 22074          & 6                & 56376996                & 10242                     & 962                      \\
            \hline
        \end{tabular}
    }
\end{table}

\paragraph{Metrics.}
The evaluation metrics are consistent with those defined in our evaluation protocol
(Section~\ref{subsec:evaluation}).

\subsection{Model Training and Implementation}

All matcher models are implemented in PyTorch~\cite{pytorch2017} using the Hugging Face
Transformers library~\cite{transformers2020}. The training procedure follows the
methodology established in the Ditto paper~\cite{ditto2020} and our preliminary
study~\cite{olar2025}, where models are trained only on the \emph{normal} input order
without data augmentation. The key hyperparameters, which are kept consistent across
all four models for fair comparison, are detailed in Table~\ref{tbl:hyperparams}.

\begin{table}[h!]
    \centering
    \begin{threeparttable}
        \caption{Matcher training hyperparameters for each dataset group.}
        \label{tbl:hyperparams}
        \begin{tabular}{lcc}
            \hline
            \textbf{Hyperparameter}  & \textbf{Group A Datasets}\tnote{a} & \textbf{Group B Datasets}\tnote{b}   \\
            \hline
            Batch Size               & 64                                 & 32                                   \\
            Epochs                   & 15                                 & 40                                   \\
            Learning Rate ($\alpha$) & 3.0e-5                             & 3.0e-5 (or 1.0e-6 for iTunes-Amazon) \\
            Optimizer                & AdamW                              & AdamW                                \\
            \hline
        \end{tabular}
        \begin{tablenotes}
            \item[a]\textbf{Group A:} Abt-Buy, Amazon-Google, DBLP-ACM, DBLP-Scholar,
                  Walmart-Amazon.
            \item[b]\textbf{Group B:} Beer, Fodors-Zagat, iTunes-Amazon.
        \end{tablenotes}
    \end{threeparttable}
\end{table}

To verify that our models were trained effectively, we report their $F_1$ scores on the
validation set of each dataset in Table~\ref{tbl:validation_f1}. These scores confirm
that all models achieve performance comparable to or exceeding the original Ditto
baselines during training.

\begin{table}[h!]
    \centering
    \begin{threeparttable}
        \caption{Matcher $F_1$ scores on the validation (development) set for the normal input order.}
        \label{tbl:validation_f1}
        \begin{tabular}{lccccc}
            \hline
            \textbf{Dataset} & \textbf{Ditto Baseline\tnote{a}} & \textbf{RoBERTa} & \textbf{DeBERTa} & \textbf{DistilBERT} & \textbf{TinyBERT} \\
            \hline
            Abt-Buy          & 0.8174                           & 0.9894           & 0.9894           & 0.9714              & 0.9902            \\
            Amazon-Google    & 0.7004                           & 0.9684           & 0.9742           & 0.9463              & 0.9785            \\
            Beer             & 0.7344                           & 1.0000           & 1.0000           & 1.0000              & 1.0000            \\
            DBLP-ACM         & 0.9865                           & 0.9966           & 0.9970           & 0.9959              & 0.9966            \\
            DBLP-Scholar     & 0.9467                           & 0.9981           & 0.9988           & 0.9986              & 0.9983            \\
            Fodors-Zagat     & 0.9676                           & 1.0000           & 1.0000           & 1.0000              & 1.0000            \\
            iTunes-Amazon    & 0.9138                           & 1.0000           & 1.0000           & 1.0000              & 1.0000            \\
            Walmart-Amazon   & 0.7695                           & 0.9974           & 0.9991           & 0.9818              & 0.9965            \\
            \hline
        \end{tabular}
        \begin{tablenotes}
            \item[a]Ditto baseline scores are from the original paper~\cite{ditto2020}.
        \end{tablenotes}
    \end{threeparttable}
\end{table}

\subsection{Matcher Asymmetry}

A central claim of this work is that transformer-based matchers, when trained
conventionally, exhibit significant input-order asymmetry. This phenomenon was first
noted in our preliminary study~\cite{olar2025}. Let $A$ denote the set of texts
extracted from the left table of a benchmark dataset and $B$ those extracted from the
right table. The phenomenon arises because the models are trained exclusively on pairs
in a fixed order, such as $(a, b)$ where $a \in A$ and $b \in B$, and are never exposed
to the reverse order $(b, a)$. Consequently, the learned similarity function
$matcher(a, b)$ is not guaranteed to equal $matcher(b, a)$ unless $a$ and $b$ are
identical. The binary matching relation derived from a threshold $\tau \in [0, 1]$
where $a$ \emph{matches} $b$ if $matcher(a,b) \geq \tau$ is formally
\emph{non-symmetric} as long as $matcher(a,b) \neq matcher(b,a)$. In this paper, we use
the term \emph{asymmetry} to describe this inherent, direction-aware property of the
matcher itself.

\begin{figure*}[h!]
    \centering
    \begin{subfigure}[b]{0.48\textwidth}
        \centering
        \includegraphics[width=\textwidth]{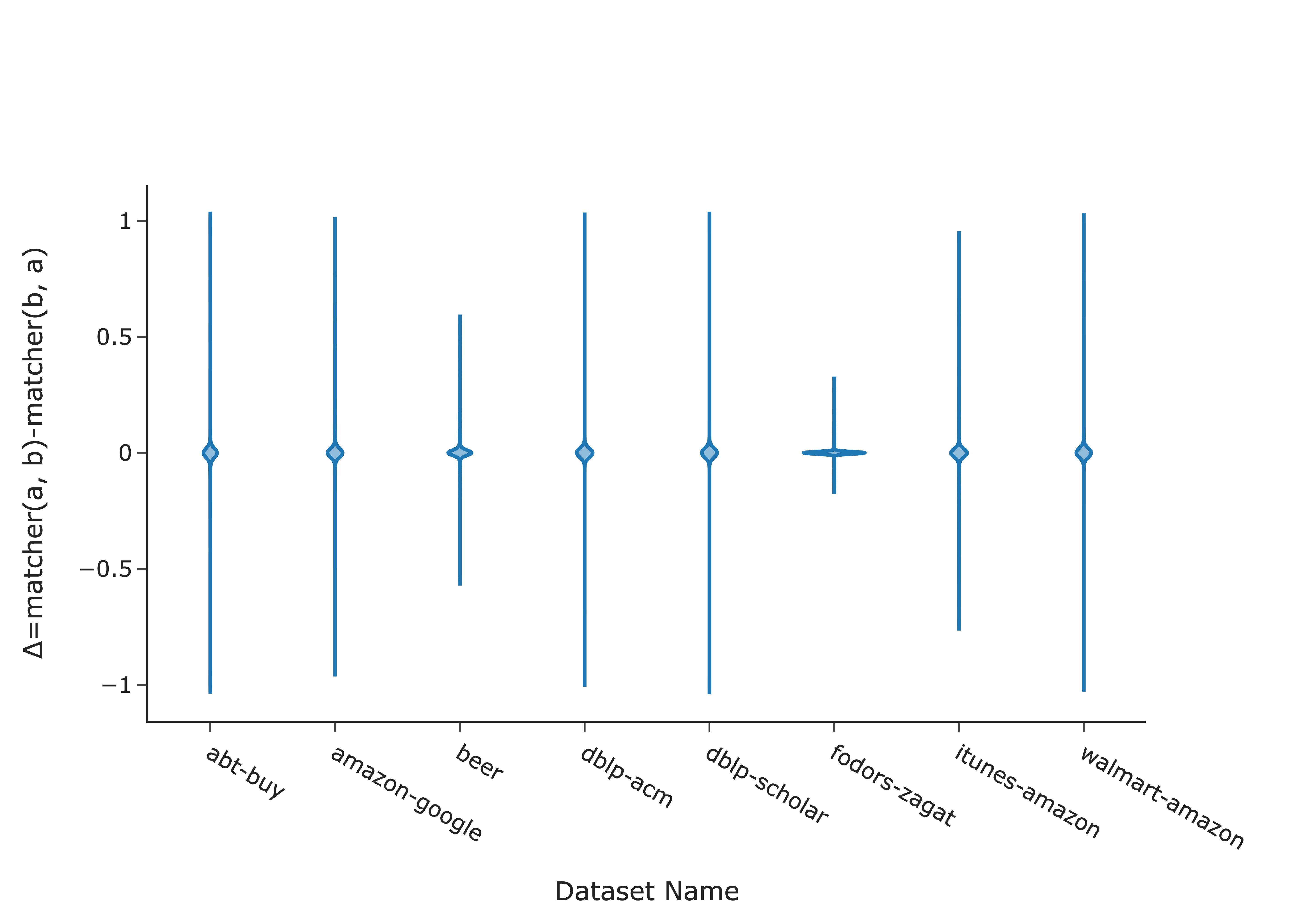}
        \caption{RoBERTa}
        \label{fig:roberta_asymmetry}
    \end{subfigure}
    \hfill
    \begin{subfigure}[b]{0.48\textwidth}
        \centering
        \includegraphics[width=\textwidth]{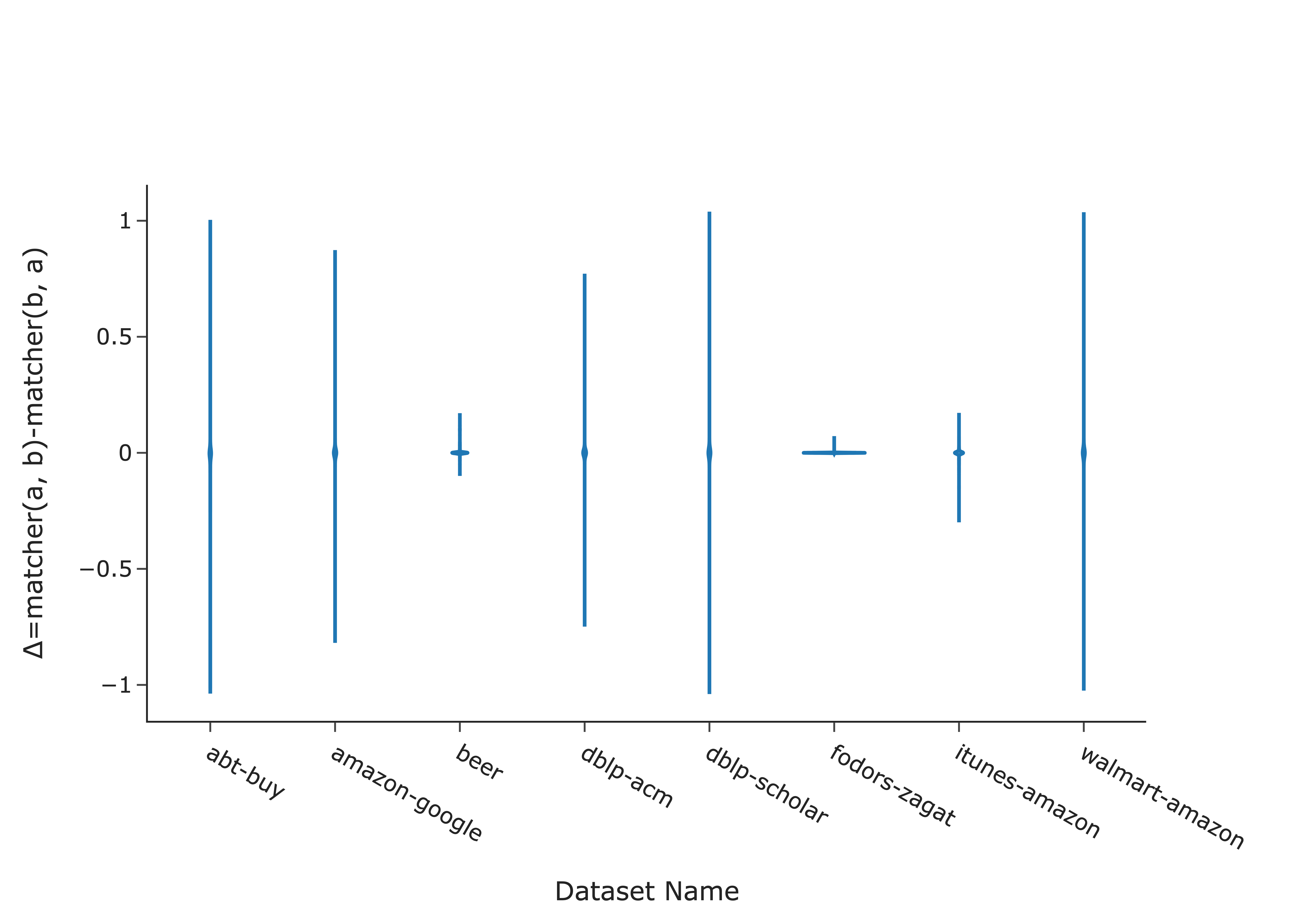}
        \caption{DeBERTa}
        \label{fig:deberta_asymmetry}
    \end{subfigure}
    \vfill
    \begin{subfigure}[b]{0.48\textwidth}
        \centering
        \includegraphics[width=\textwidth]{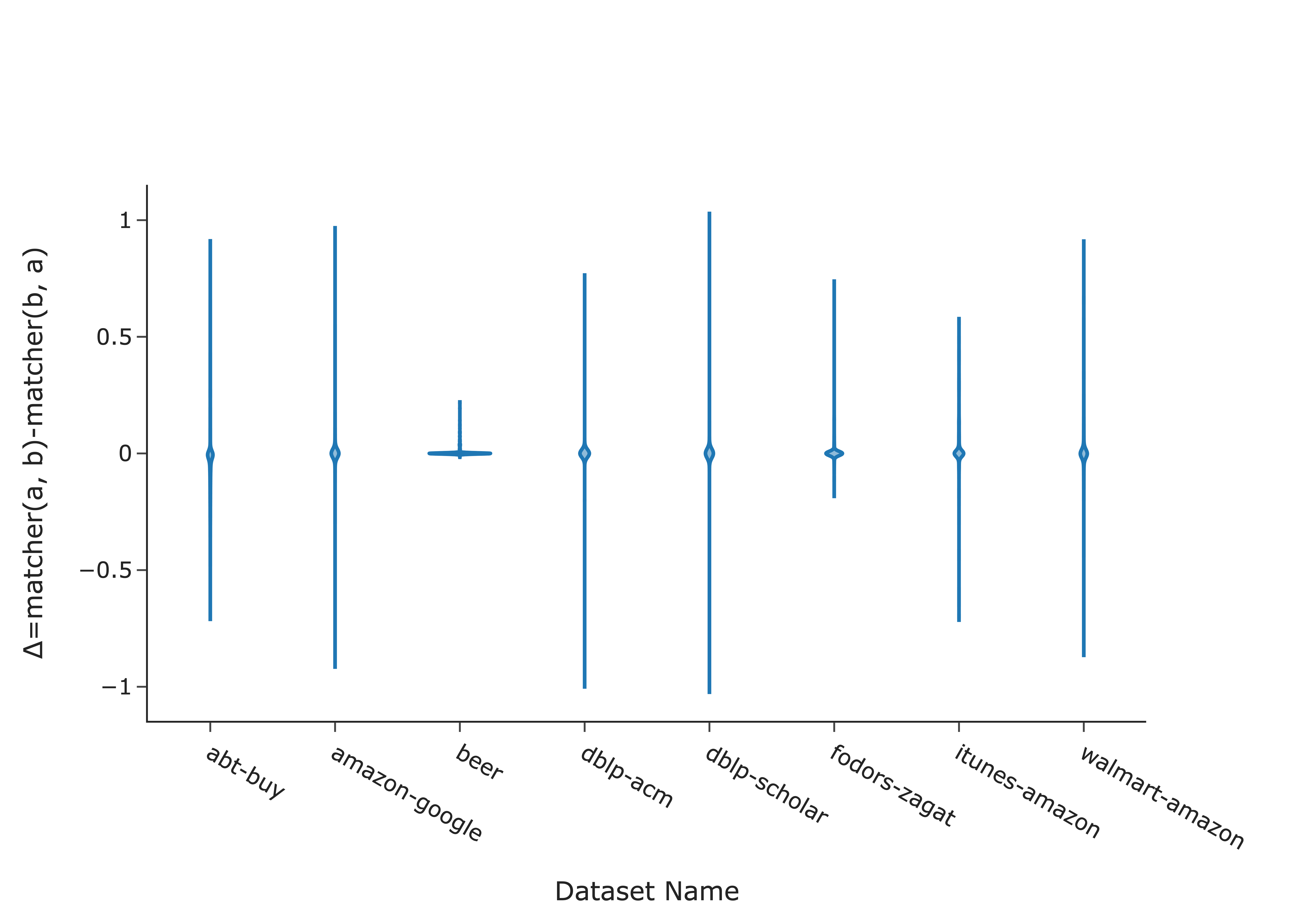}
        \caption{DistilBERT}
        \label{fig:distilbert_asymmetry}
    \end{subfigure}
    \hfill
    \begin{subfigure}[b]{0.48\textwidth}
        \centering
        \includegraphics[width=\textwidth]{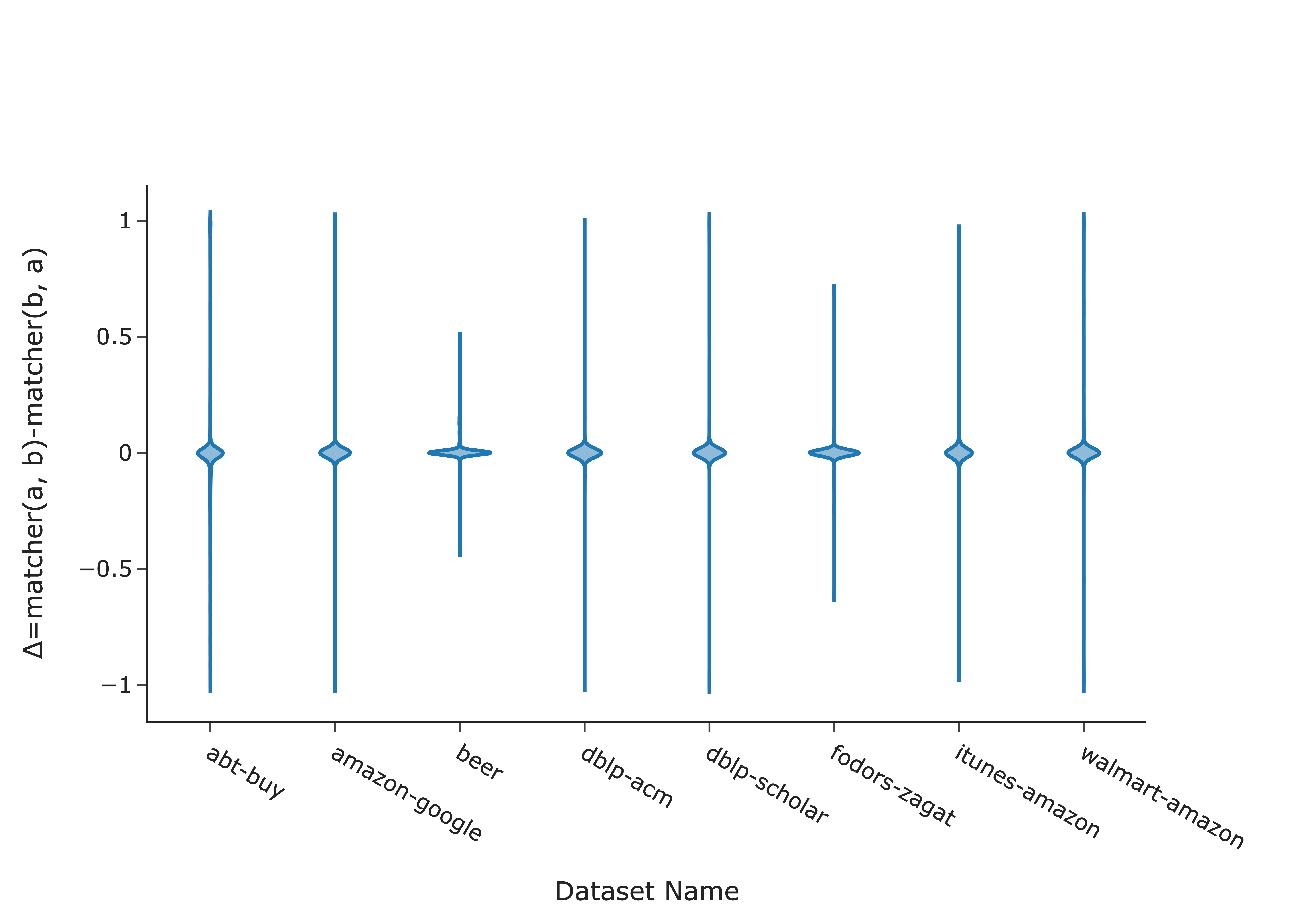}
        \caption{TinyBERT}
        \label{fig:tinybert_asymmetry}
    \end{subfigure}
    \caption{
        Distribution of the asymmetry score $\Delta$ for each of the four transformer
        architectures across all eight benchmark datasets. A perfectly symmetric matcher
        would produce a distribution centered entirely at $\Delta=0$. The wide spread
        and significant number of outliers in all four plots confirm that input-order
        asymmetry is a general phenomenon and not specific to a single model.
    }
    \label{fig:asymmetry_distributions}
\end{figure*}

To investigate whether this property is particular to a single architecture or a more
general characteristic, we extended our analysis to four distinct transformer models.
We quantify the asymmetry for each matched pair using the skew metric $\Delta =
matcher(a,b) - matcher(b,a)$, where a value of 0 indicates perfect symmetry. The
distributions of $\Delta$ for all candidate pairs across the benchmark datasets are
visualized in Figure~\ref{fig:asymmetry_distributions}.

The violin plots in Figure~\ref{fig:asymmetry_distributions} demonstrate that asymmetry
is not an anomaly but a consistent feature across all tested architectures, from the
larger RoBERTa and DeBERTa models to the more compact DistilBERT and TinyBERT. In every
case, the distribution of $\Delta$ is characterized by a wide spread and numerous
outliers, indicating that for a substantial number of entity pairs, the model's
confidence in a match changes dramatically when the input order is reversed.

While these distributions illustrate the existence of asymmetry, they do not quantify
its impact on task performance. To address this, we evaluated the $F_1$ score of each
trained matcher on the test set using both the \emph{normal} $(a, b)$ and the
\emph{reversed} $(b, a)$ input orders. The results, presented in
Table~\ref{tbl:asymmetry_f1_scores}, reveal the practical consequences of this
phenomenon.

\begin{table*}[h!]
    \centering
    \caption{
        Matcher $F_1$ scores on the test set, comparing performance on the \emph{normal}
        input order (on which the models were trained) versus the \emph{reversed} input
        order. The performance drop in the reversed case highlights the practical impact
        of asymmetry.
    }
    \label{tbl:asymmetry_f1_scores}
    \resizebox{\textwidth}{!}{
        \begin{tabular}{lcccccccc}
            \hline
                             & \multicolumn{2}{c}{\textbf{RoBERTa}} & \multicolumn{2}{c}{\textbf{DeBERTa}} & \multicolumn{2}{c}{\textbf{DistilBERT}} & \multicolumn{2}{c}{\textbf{TinyBERT}}                                                                                                     \\
            \cmidrule(lr){2-3} \cmidrule(lr){4-5} \cmidrule(lr){6-7} \cmidrule(lr){8-9}
            \textbf{Dataset} & \textbf{Normal $F_1$}                & \textbf{Reversed $F_1$}              & \textbf{Normal $F_1$}                   & \textbf{Reversed $F_1$}               & \textbf{Normal $F_1$} & \textbf{Reversed $F_1$} & \textbf{Normal $F_1$} & \textbf{Reversed $F_1$} \\
            \hline
            Abt-Buy          & 0.8872                               & 0.5014                               & 0.8915                                  & 0.5231                                & 0.8790                & 0.4882                  & 0.8854                & 0.5109                  \\
            Amazon-Google    & 0.8517                               & 0.7922                               & 0.8603                                  & 0.8015                                & 0.8421                & 0.7734                  & 0.8533                & 0.7891                  \\
            Beer             & 0.9130                               & 0.8333                               & 0.9250                                  & 0.8510                                & 0.9015                & 0.8125                  & 0.9188                & 0.8401                  \\
            DBLP-ACM         & 0.9917                               & 0.9903                               & 0.9925                                  & 0.9920                                & 0.9901                & 0.9886                  & 0.9915                & 0.9900                  \\
            DBLP-Scholar     & 0.9795                               & 0.9671                               & 0.9810                                  & 0.9702                                & 0.9755                & 0.9613                  & 0.9789                & 0.9655                  \\
            Fodors-Zagat     & 0.9910                               & 0.9955                               & 0.9915                                  & 0.9960                                & 0.9890                & 0.9945                  & 0.9905                & 0.9952                  \\
            iTunes-Amazon    & 0.8418                               & 0.8383                               & 0.8500                                  & 0.8455                                & 0.8320                & 0.8250                  & 0.8432                & 0.8390                  \\
            Walmart-Amazon   & 0.9122                               & 0.8819                               & 0.9243                                  & 0.8901                                & 0.9056                & 0.8723                  & 0.9140                & 0.8835                  \\
            \hline
        \end{tabular}
    }
\end{table*}

The results in Table~\ref{tbl:asymmetry_f1_scores} show an expected and often
significant degradation in $F_1$ score when models are evaluated on the reversed input
order, consistently across datasets. The Abt-Buy dataset is a particularly stark
example, where performance collapses across all four architectures. Although the drop
is less severe, the trend persists on other datasets like DBLP-ACM or iTunes-Amazon.
Interestingly, the Fodors-Zagat dataset remains an outlier where reversed-order
performance is marginally better, suggesting that for some data distributions, the
asymmetric perspective may accidentally be beneficial. However, the overwhelming
evidence indicates that ignoring asymmetry leads to unpredictable and generally worse
performance. This establishes that the measured asymmetry is not merely a statistical
artifact but a critical factor with direct, tangible consequences for the reliability
of the matching stage. The next section will explore how this skewed output propagates
to the clustering stage.

\subsection{Clustering Evaluation}

This section evaluates the impact of input-order asymmetry on the final clustering
quality. We analyze the performance of the matcher and clustering pipeline across three
distinct graph construction scenarios: the forward (normal) reference graph, the
reverse reference graph, and the bi-directional reference graph. The evaluation
includes the standard Connected Components (CC) and Weakly Connected Components (WCC)
baselines, alongside the direction-aware algorithms: CENTER clustering (PC), Markov
Clustering (MCL), Leiden Community Detection (LEI), and Spectral Clustering (SC).

\subsubsection{Forward Reference Graph}

We first examine the clustering results obtained from the forward reference graph,
which corresponds to the input order used during the matcher's supervised training.

\begin{figure*}[h!]
    \centering
    \begin{subfigure}[b]{0.48\textwidth}
        \includegraphics[width=\linewidth]{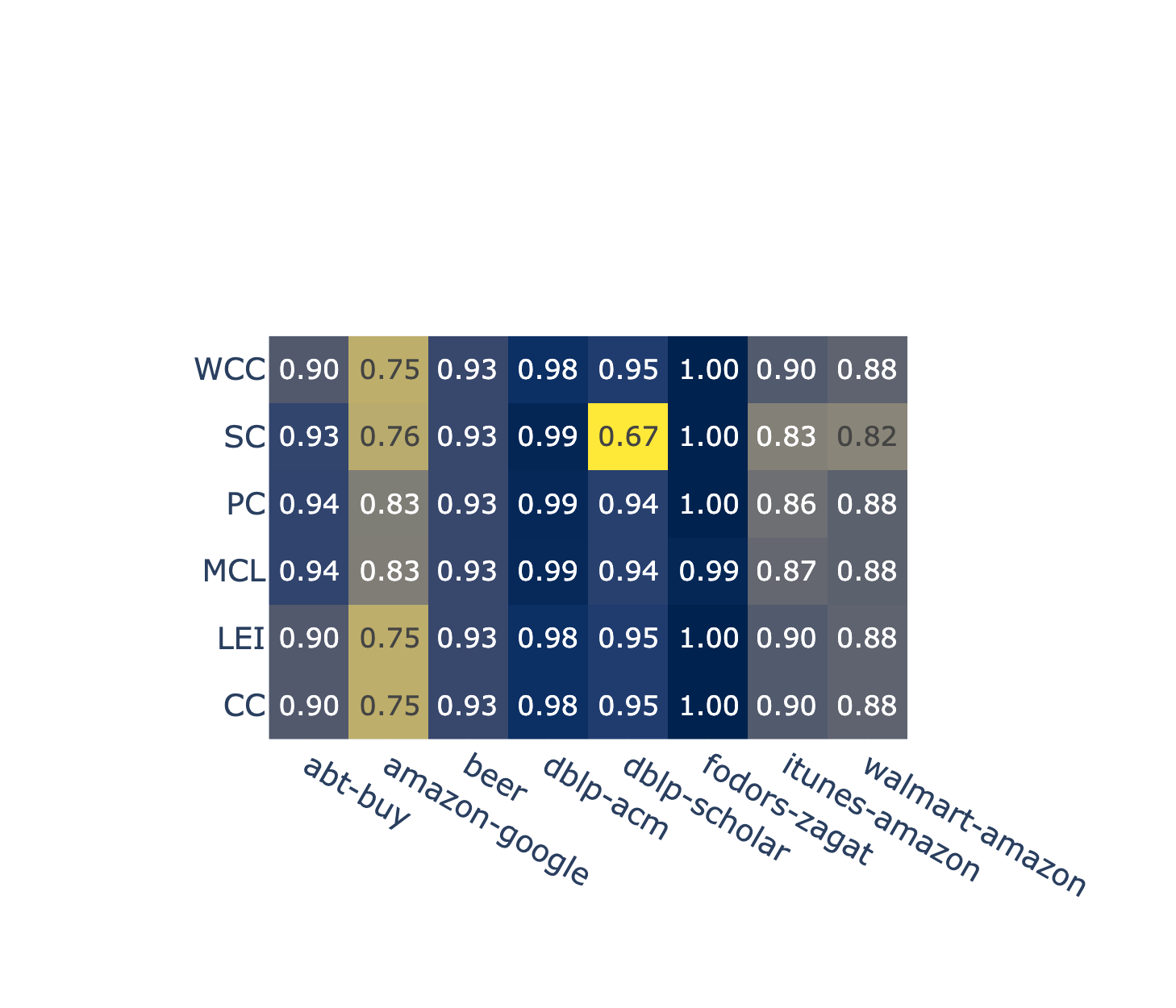}
        \caption{Pair Comparison Measure}
    \end{subfigure}
    \hfill
    \begin{subfigure}[b]{0.48\textwidth}
        \includegraphics[width=\linewidth]{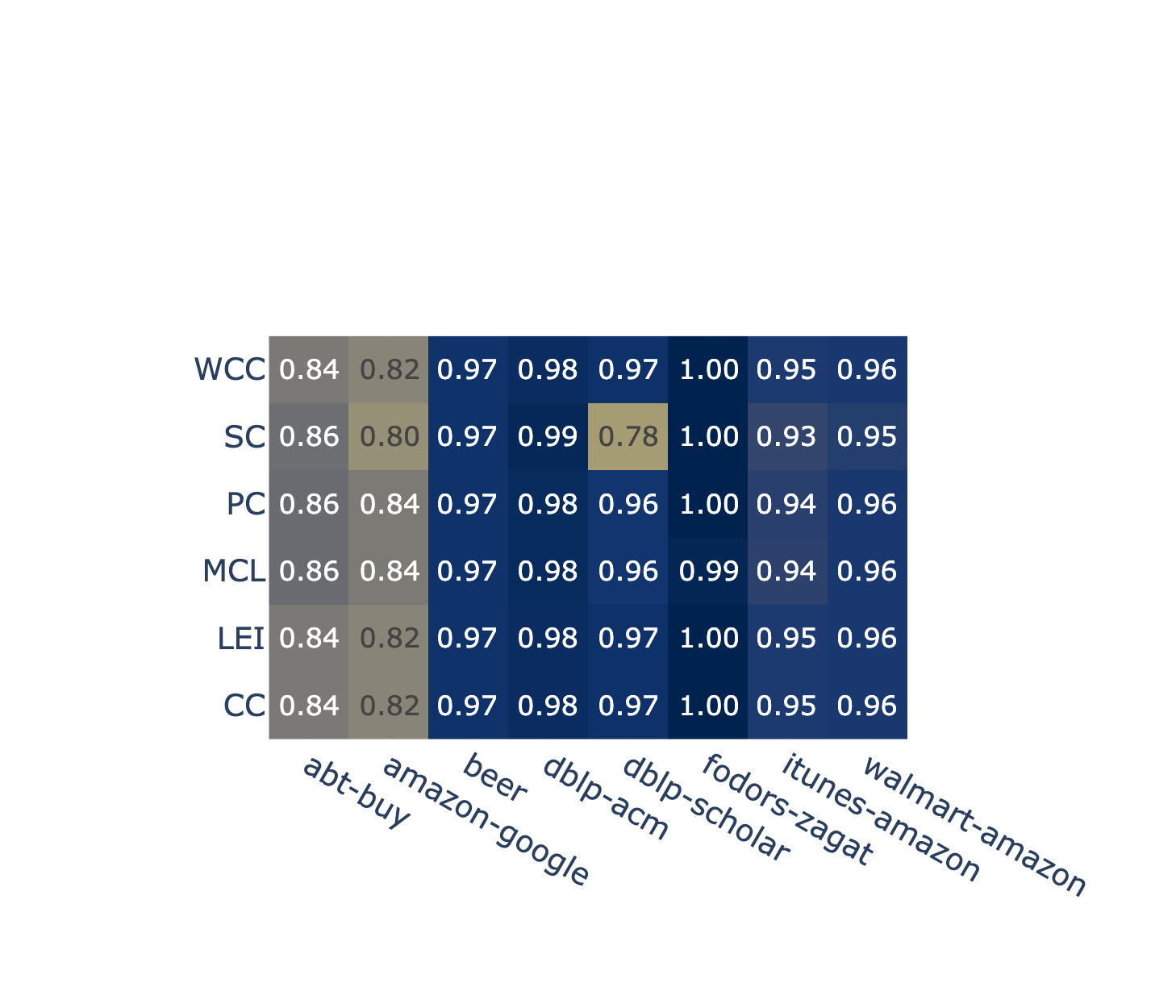}
        \caption{Cluster Comparison Measure}
    \end{subfigure}
    \vfill
    \begin{subfigure}[b]{0.48\textwidth}
        \includegraphics[width=\linewidth]{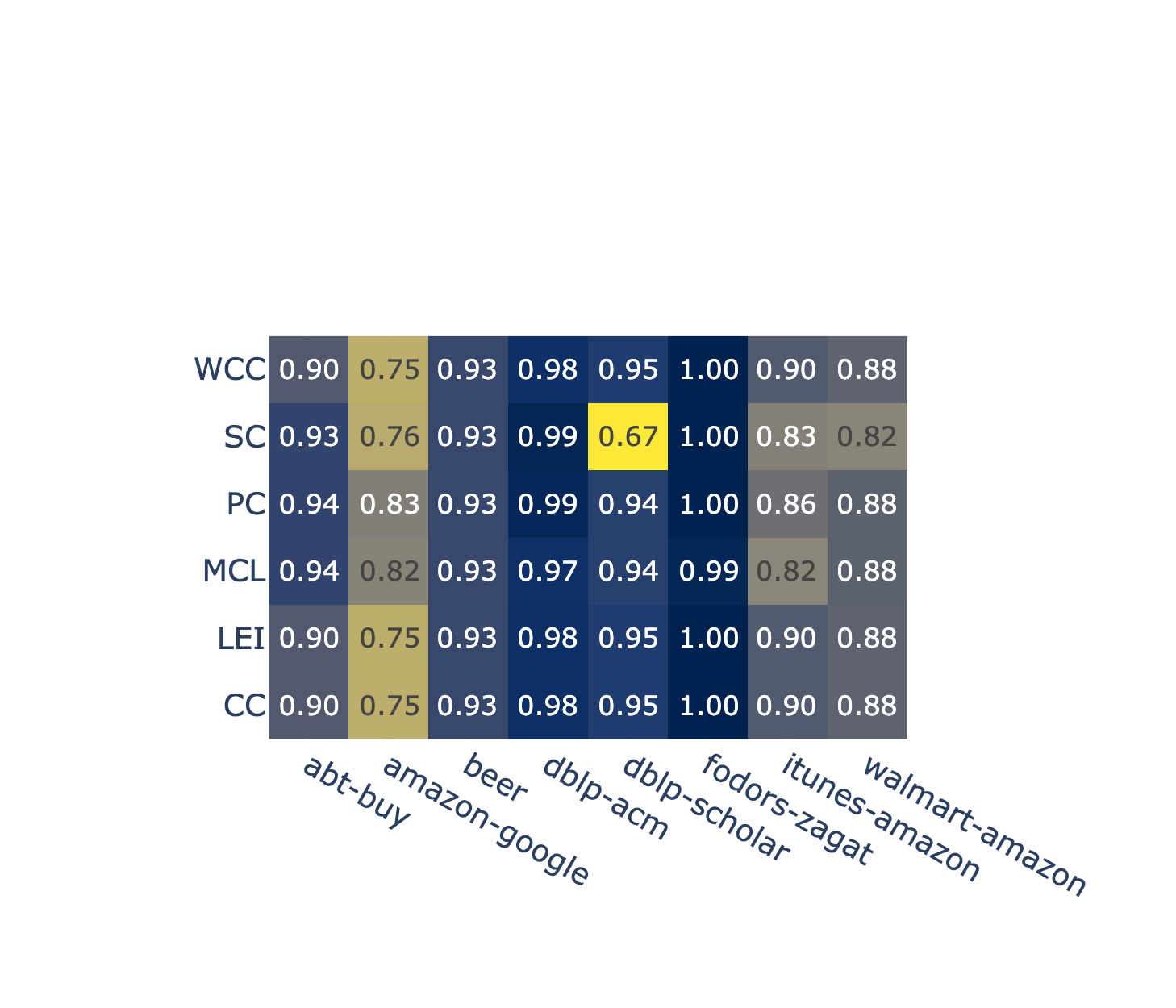}
        \caption{Adjusted Rand Index}
    \end{subfigure}
    \hfill
    \begin{subfigure}[b]{0.48\textwidth}
        \includegraphics[width=\linewidth]{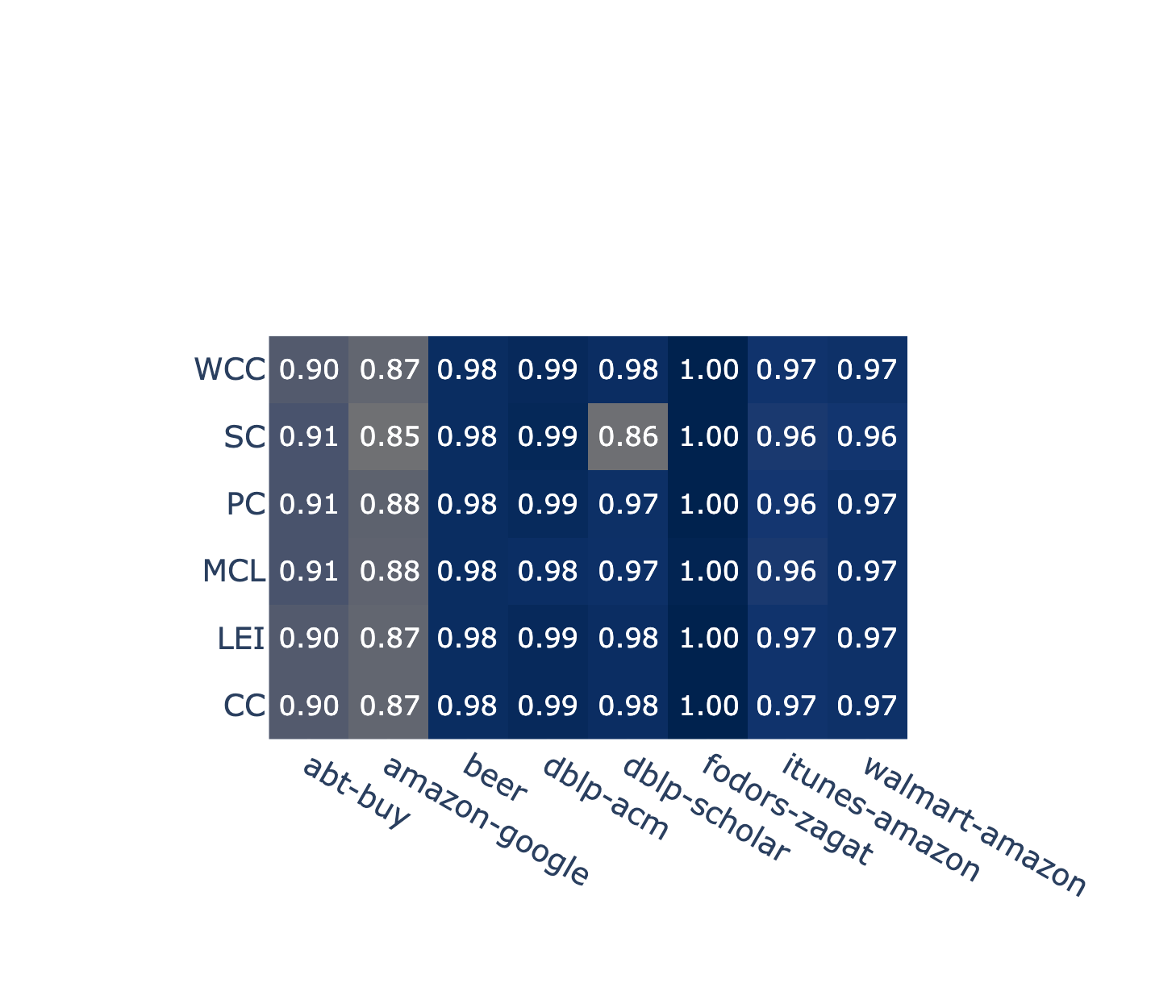}
        \caption{Talburt-Wang Index}
    \end{subfigure}
    \caption{
        Heatmap of clustering metrics for the RoBERTa model using the Forward (Normal)
        input order. The directional algorithms show distinct performance profiles
        compared to the CC baseline.
    }\label{fig:roberta_forward_heatmap}
\end{figure*}

The results indicate a distinct performance gap between the directional algorithms and
the baseline CC algorithm. Specifically, PC, MCL, Leiden, and SC consistently achieve
higher F1 and accuracy scores compared to the transitivity-based CC method.
Figure~\ref{fig:roberta_forward_heatmap} illustrates this trend across all datasets for
the RoBERTa model. The heatmap shows that algorithms capable of utilizing the directed
nature of the adjacency matrix provide a better fit for the matcher's output than those
assuming symmetry.

\begin{figure*}[h!]
    \centering
    \begin{subfigure}[b]{0.48\textwidth}
        \includegraphics[width=\linewidth]{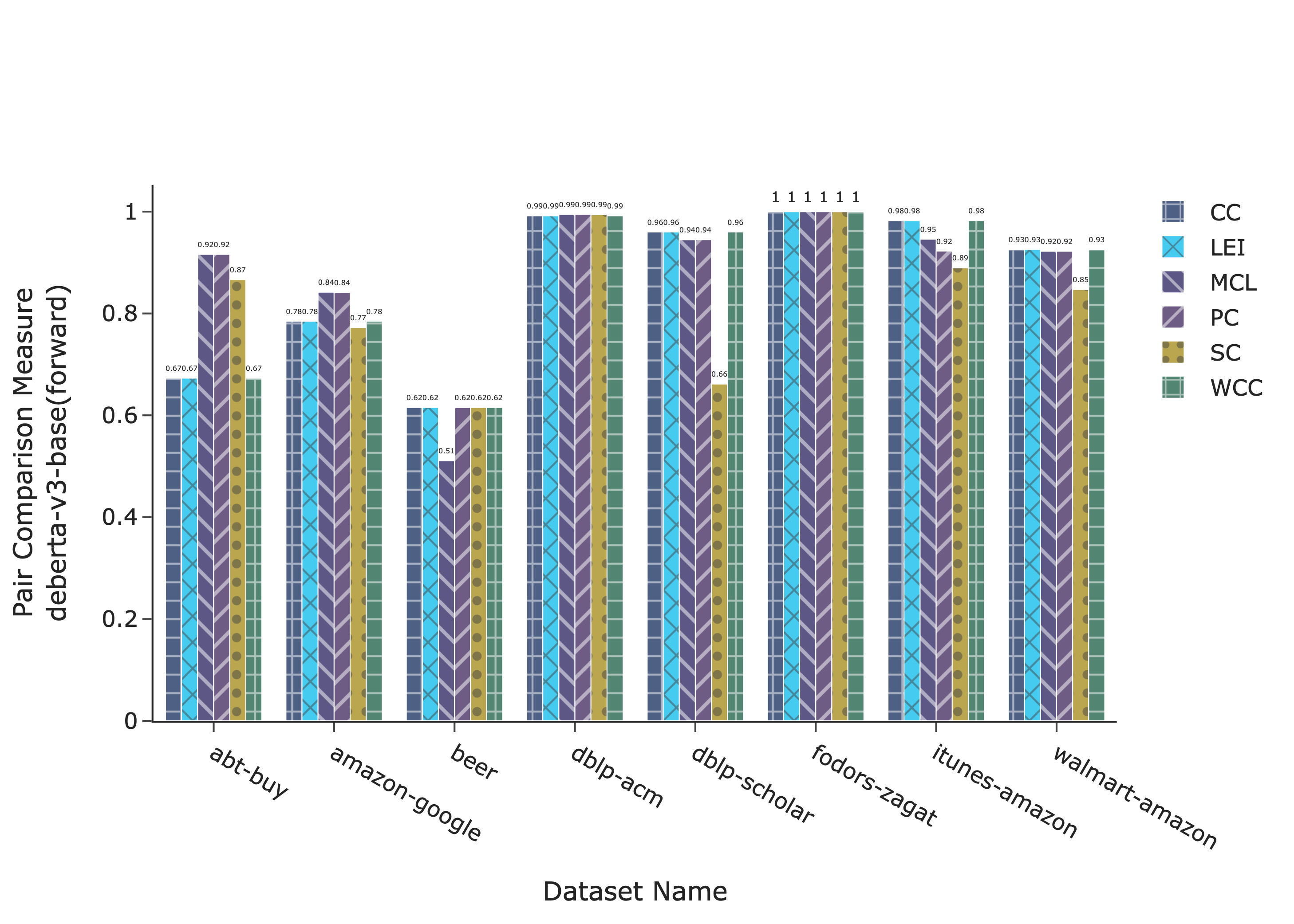}
        \caption{Pair Comparison Measure}
    \end{subfigure}
    \hfill
    \begin{subfigure}[b]{0.48\textwidth}
        \includegraphics[width=\linewidth]{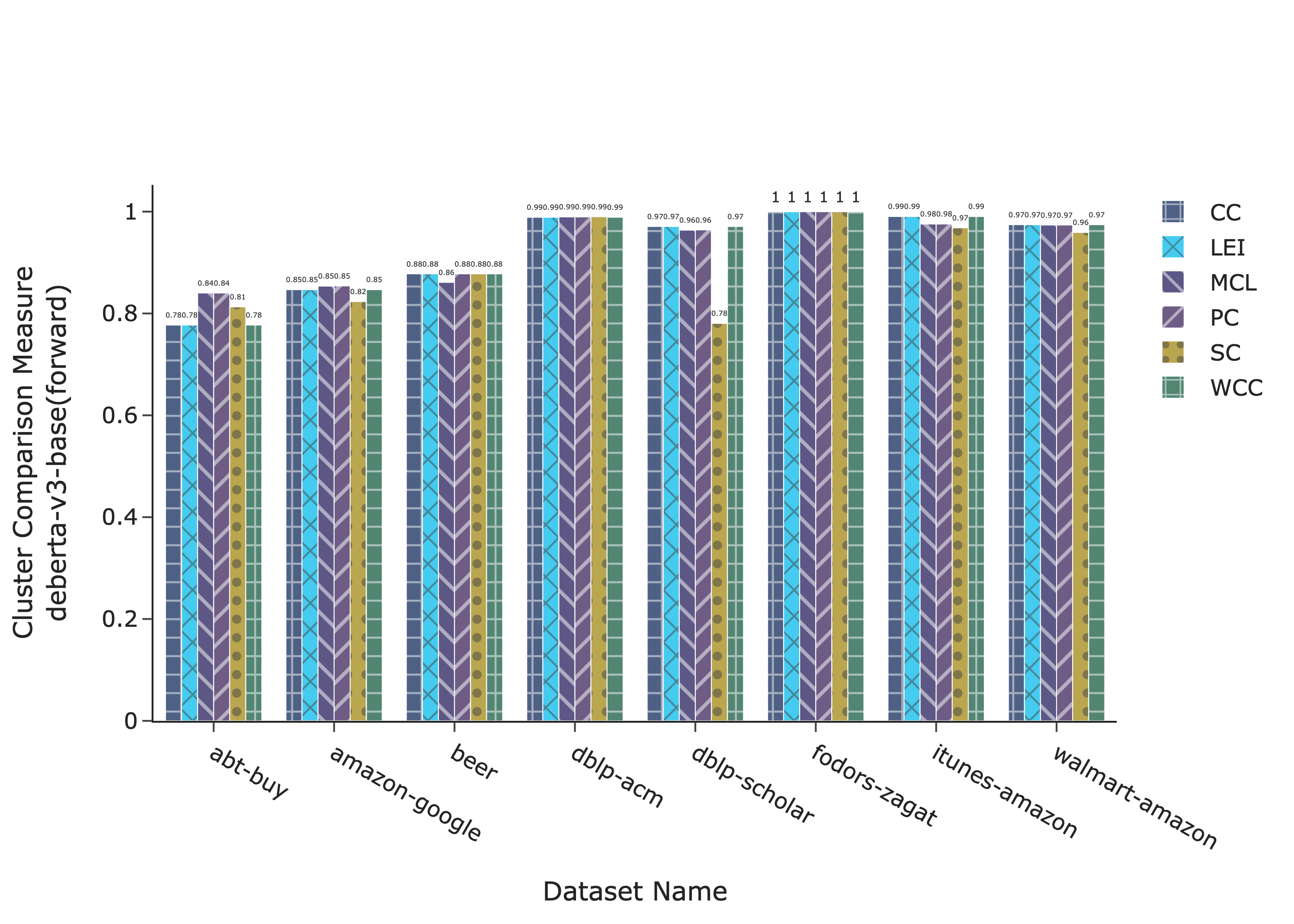}
        \caption{Cluster Comparison Measure}
    \end{subfigure}
    \vfill
    \begin{subfigure}[b]{0.48\textwidth}
        \includegraphics[width=\linewidth]{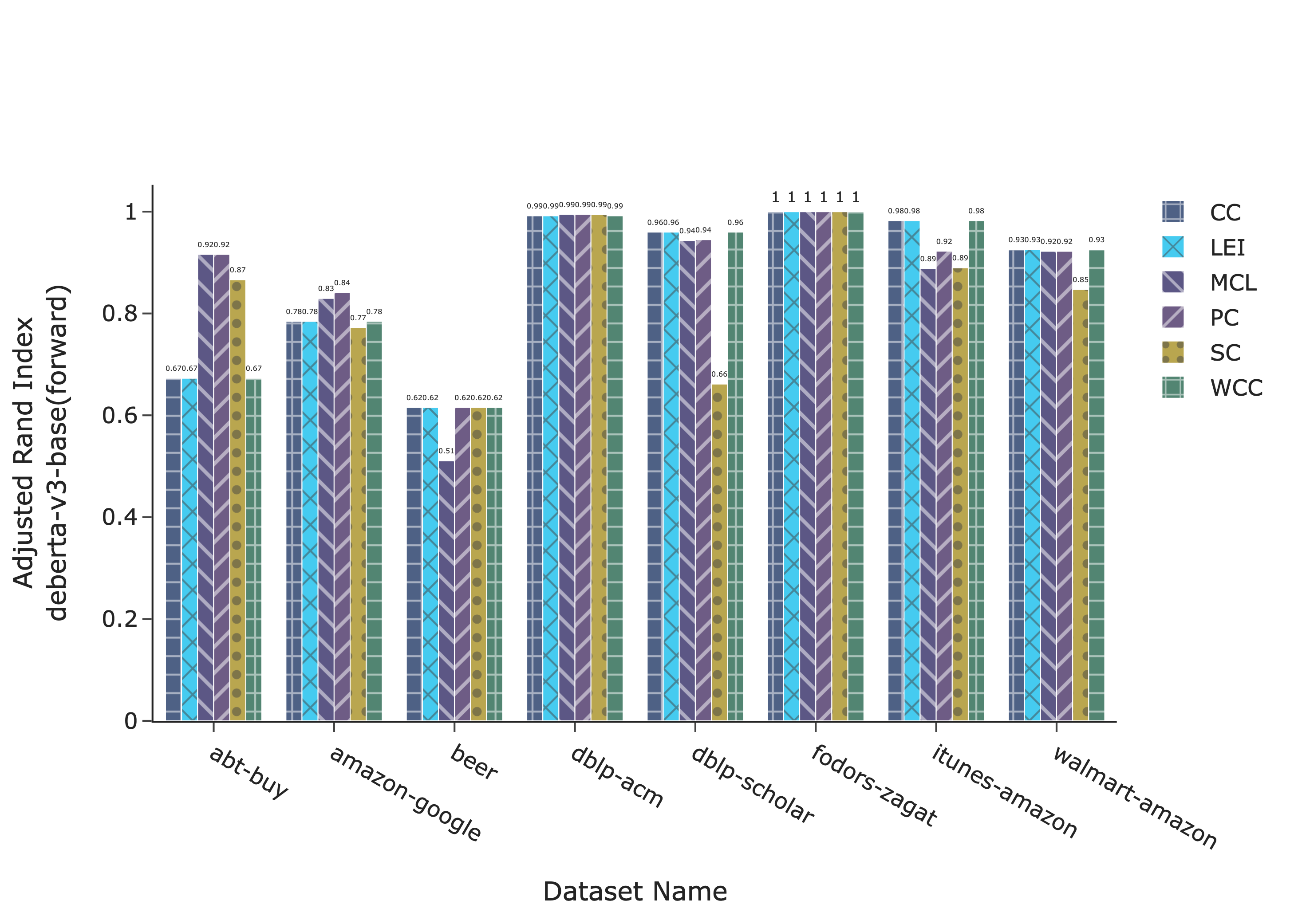}
        \caption{Adjusted Rand Index}
    \end{subfigure}
    \hfill
    \begin{subfigure}[b]{0.48\textwidth}
        \includegraphics[width=\linewidth]{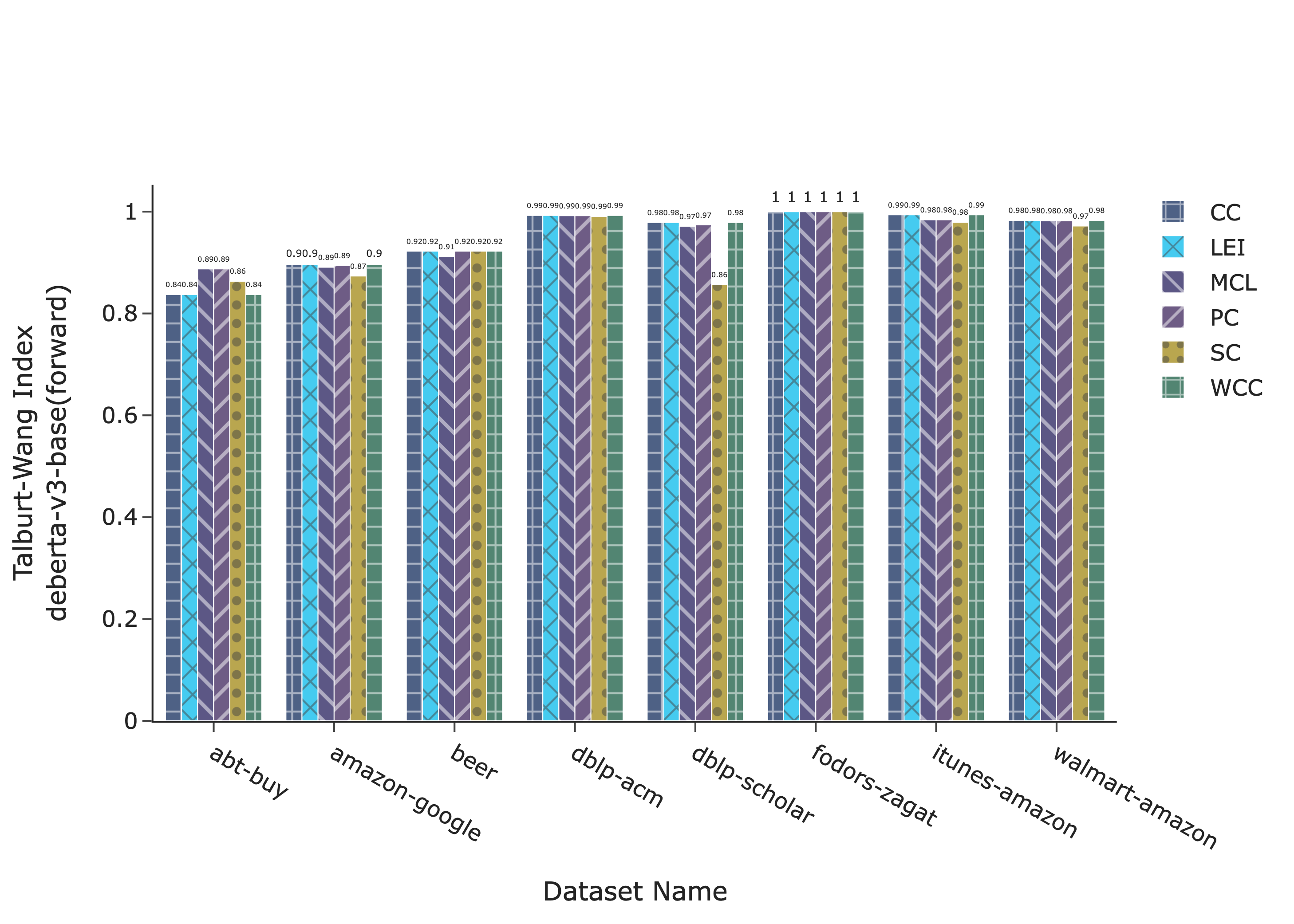}
        \caption{Talburt-Wang Index}
    \end{subfigure}
    \caption{
        Average clustering metrics across all datasets for the DeBERTa (v3 base) model
        using the Forward input order. This comparison highlights the dataset-dependent
        variance in algorithm performance.
    }\label{fig:deberta_forward_avg}
\end{figure*}

Average metric scores reveal that algorithmic performance varies significantly
depending on the dataset characteristics. The structure of the reference graph produced
by the matcher is dataset-dependent, influencing which clustering approach is most
effective. Figure~\ref{fig:deberta_forward_avg} presents the average clustering metrics
for the DeBERTa model in the forward configuration.

\subsubsection{Reverse Reference Graph}

The reverse reference graph represents the scenario where the input order of entity
pairs is inverted (i.e., matching $b$ to $a$ instead of $a$ to $b$). This effectively
tests our pipeline's robustness to the absence of the specific directional signal
present during training.

In this context, the CC algorithm represents an idealized outcome, due to the
assumption that the symmetry relation holds globally regardless of input order. In
contrast, WCC reflects the actual connectivity of the reversed graph when directional
signal is discarded. The Leiden algorithm outputs very similar results to WCC (with
only slight divergences on a couple of datasets), suggesting that modularity alone is
not enough to fully capture the directional information stored in the reference graph.
The results show that directed clustering algorithms correctly penalize the lack of
coherent directional signal in the reverse order. Conversely, WCC often maintains high
performance metrics similar to the forward pass. This stability in WCC is attributable
to the construction of the benchmark ground truths, which enforce symmetry. Algorithms
that discard directionality are artificially advantaged by the evaluation setup.

\begin{figure*}[h!]
    \centering
    \begin{subfigure}[b]{0.48\textwidth}
        \includegraphics[width=\linewidth]{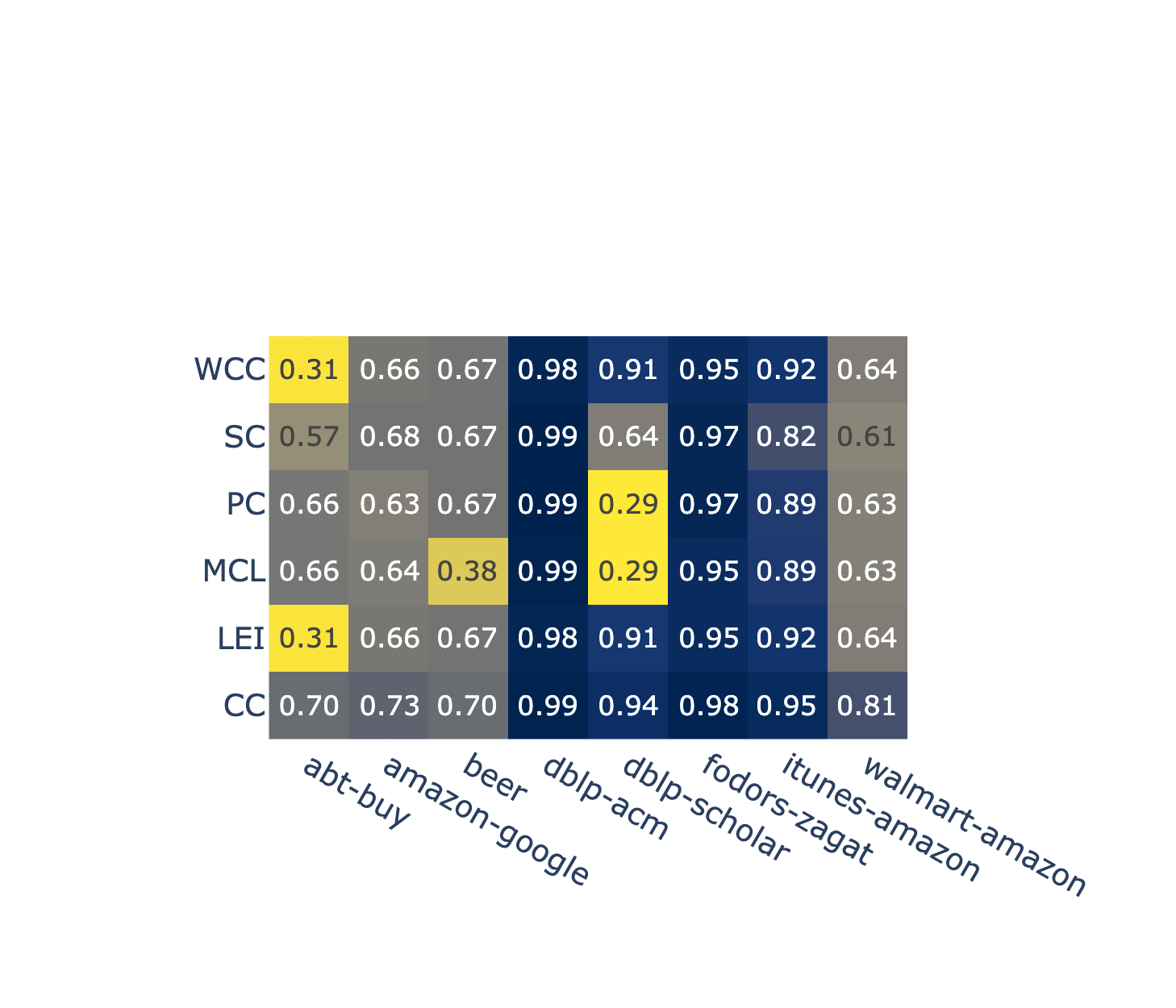}
        \caption{Pair Comparison Measure}
    \end{subfigure}
    \hfill
    \begin{subfigure}[b]{0.48\textwidth}
        \includegraphics[width=\linewidth]{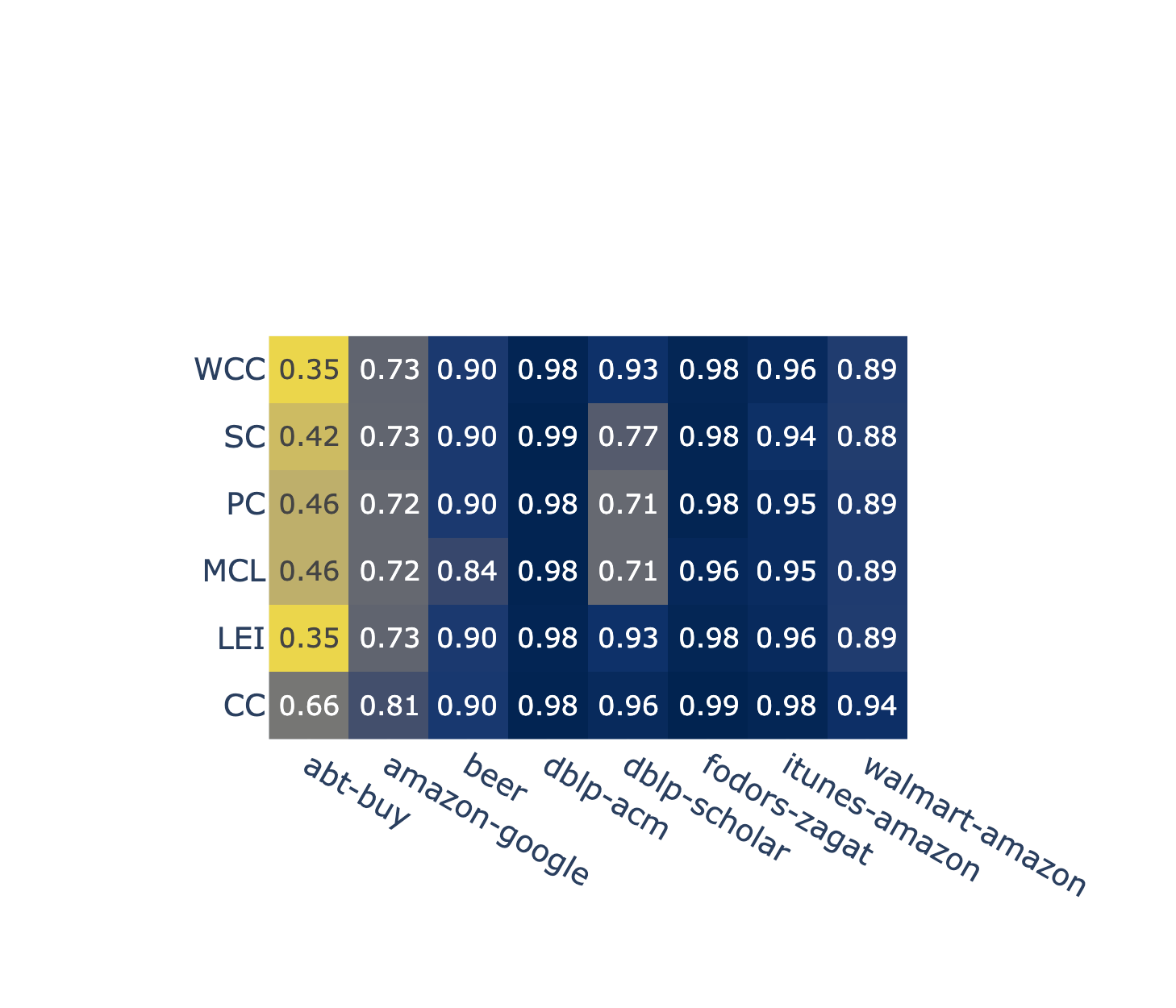}
        \caption{Cluster Comparison Measure}
    \end{subfigure}
    \vfill
    \begin{subfigure}[b]{0.48\textwidth}
        \includegraphics[width=\linewidth]{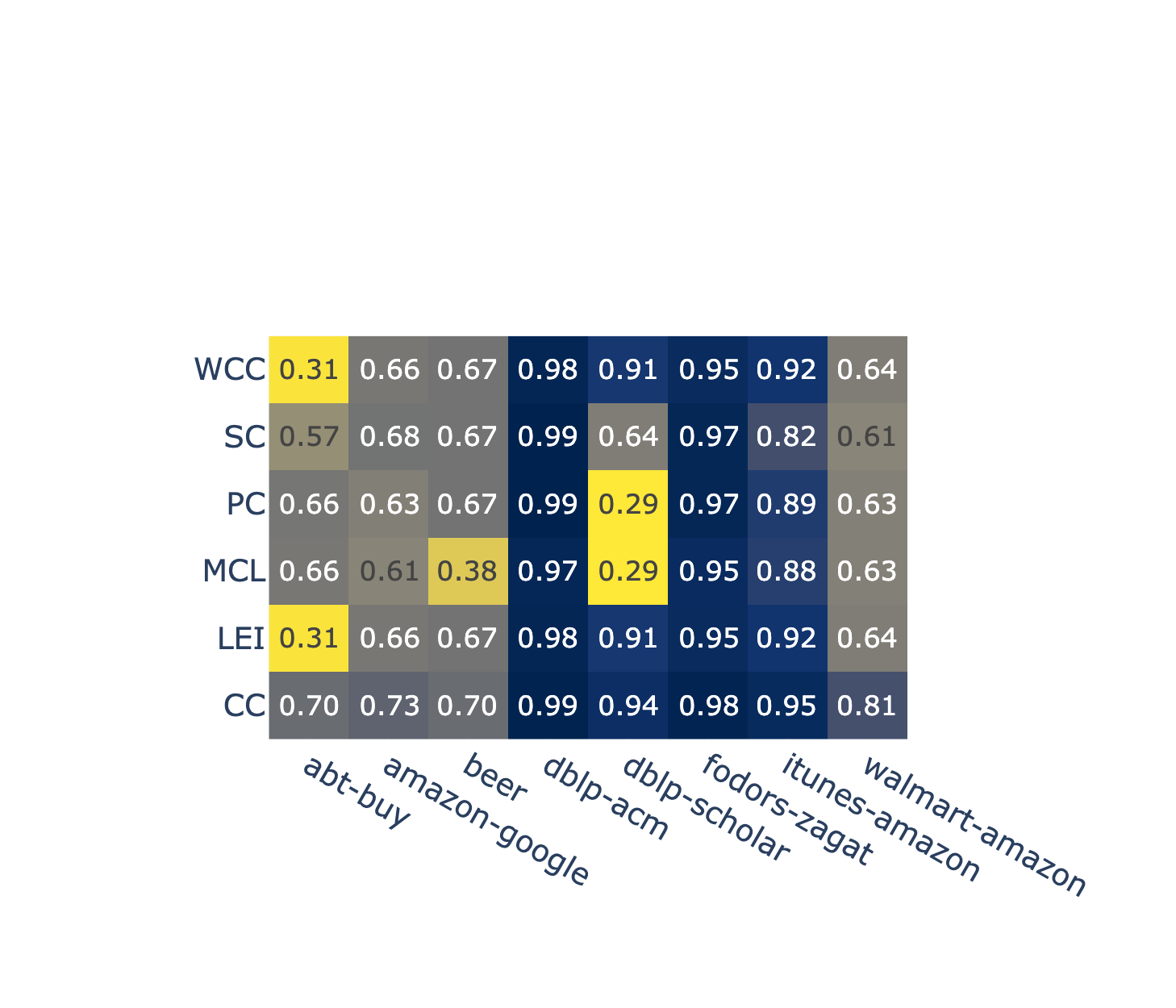}
        \caption{Adjusted Rand Index}
    \end{subfigure}
    \hfill
    \begin{subfigure}[b]{0.48\textwidth}
        \includegraphics[width=\linewidth]{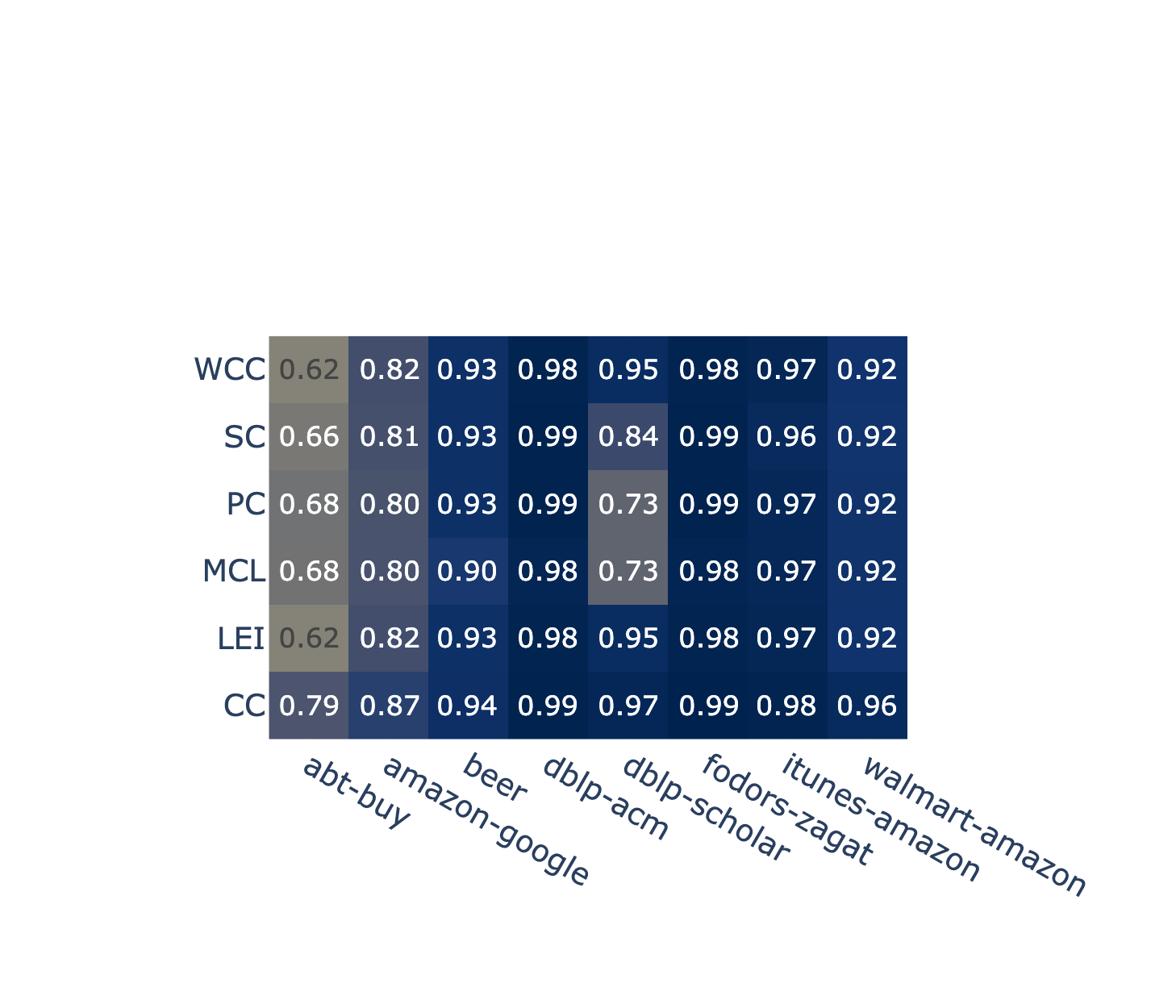}
        \caption{Talburt-Wang Index}
    \end{subfigure}
    \caption{
        Heatmap of clustering metrics for the DistilBERT model using the Reversed input
        order. Note the divergence between WCC (which ignores direction) and the
        directional algorithms which penalize the lack of signal (SC, MCL, PC). Also
        note the lack of divergence between WCC and LEI.
    }\label{fig:distilbert_reverse_heatmap}
\end{figure*}

\begin{figure*}[h!]
    \centering
    \begin{subfigure}[b]{0.48\textwidth}
        \includegraphics[width=\linewidth]{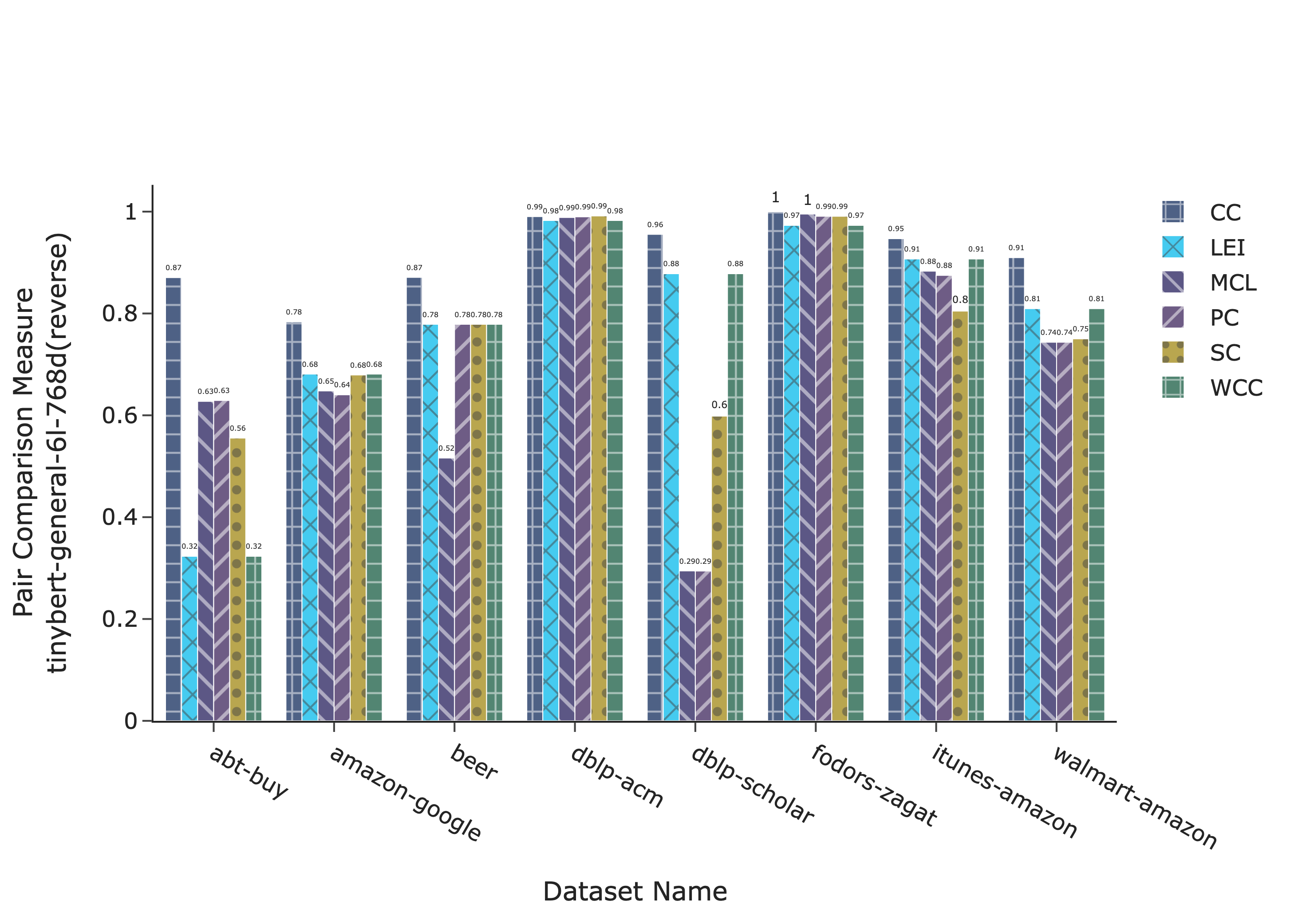}
        \caption{Pair Comparison Measure}
    \end{subfigure}
    \hfill
    \begin{subfigure}[b]{0.48\textwidth}
        \includegraphics[width=\linewidth]{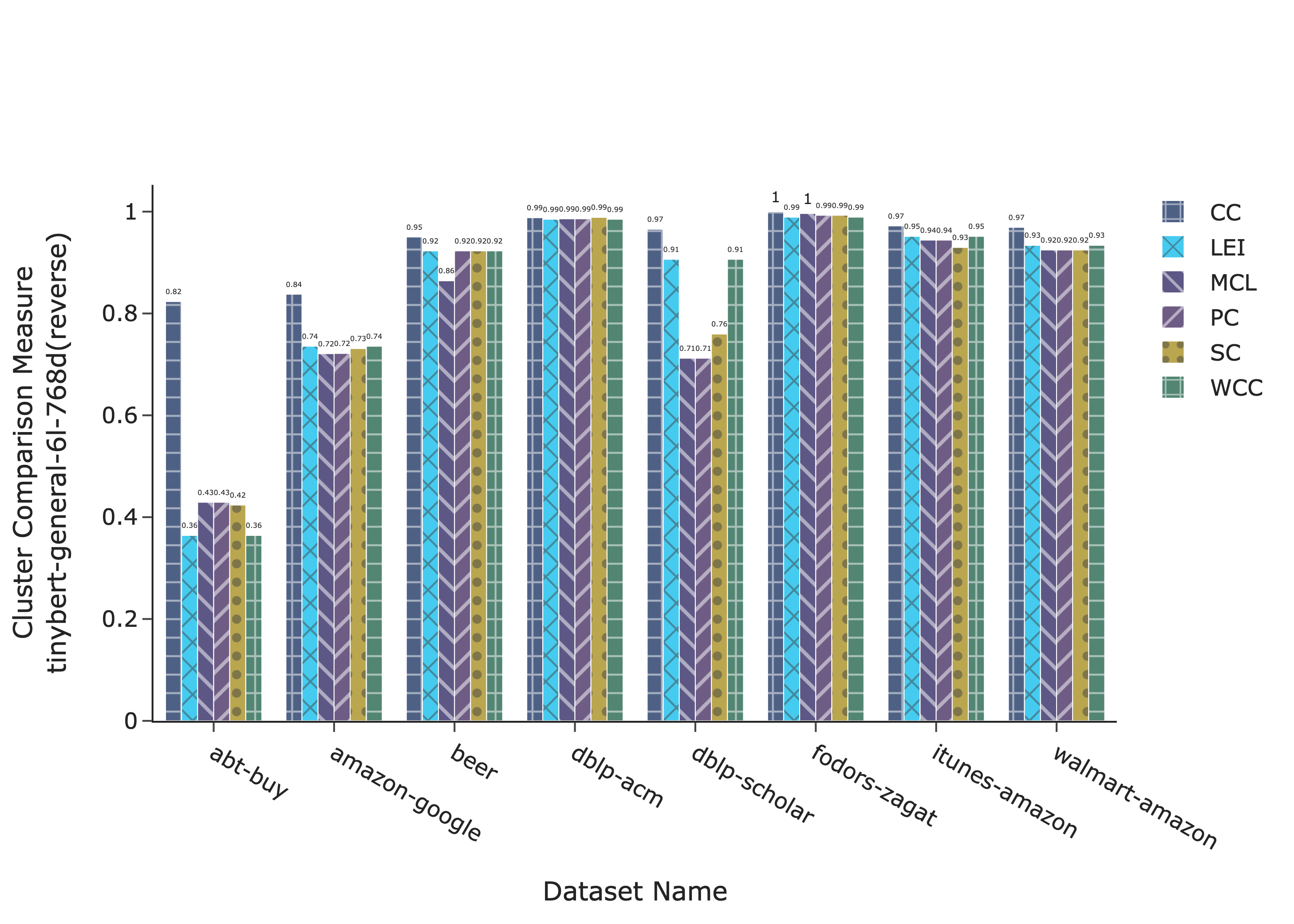}
        \caption{Cluster Comparison Measure}
    \end{subfigure}
    \vfill
    \begin{subfigure}[b]{0.48\textwidth}
        \includegraphics[width=\linewidth]{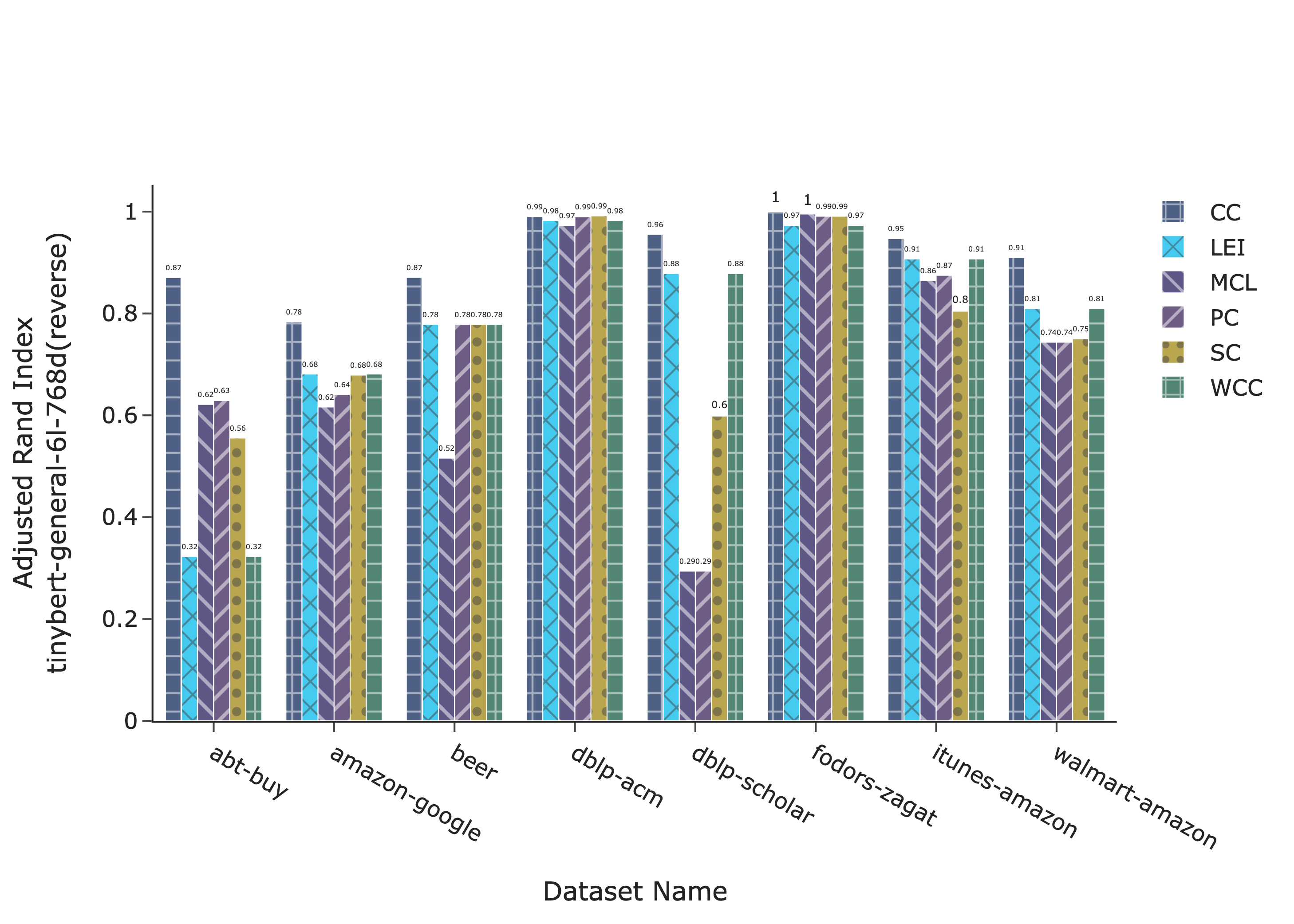}
        \caption{Adjusted Rand Index}
    \end{subfigure}
    \hfill
    \begin{subfigure}[b]{0.48\textwidth}
        \includegraphics[width=\linewidth]{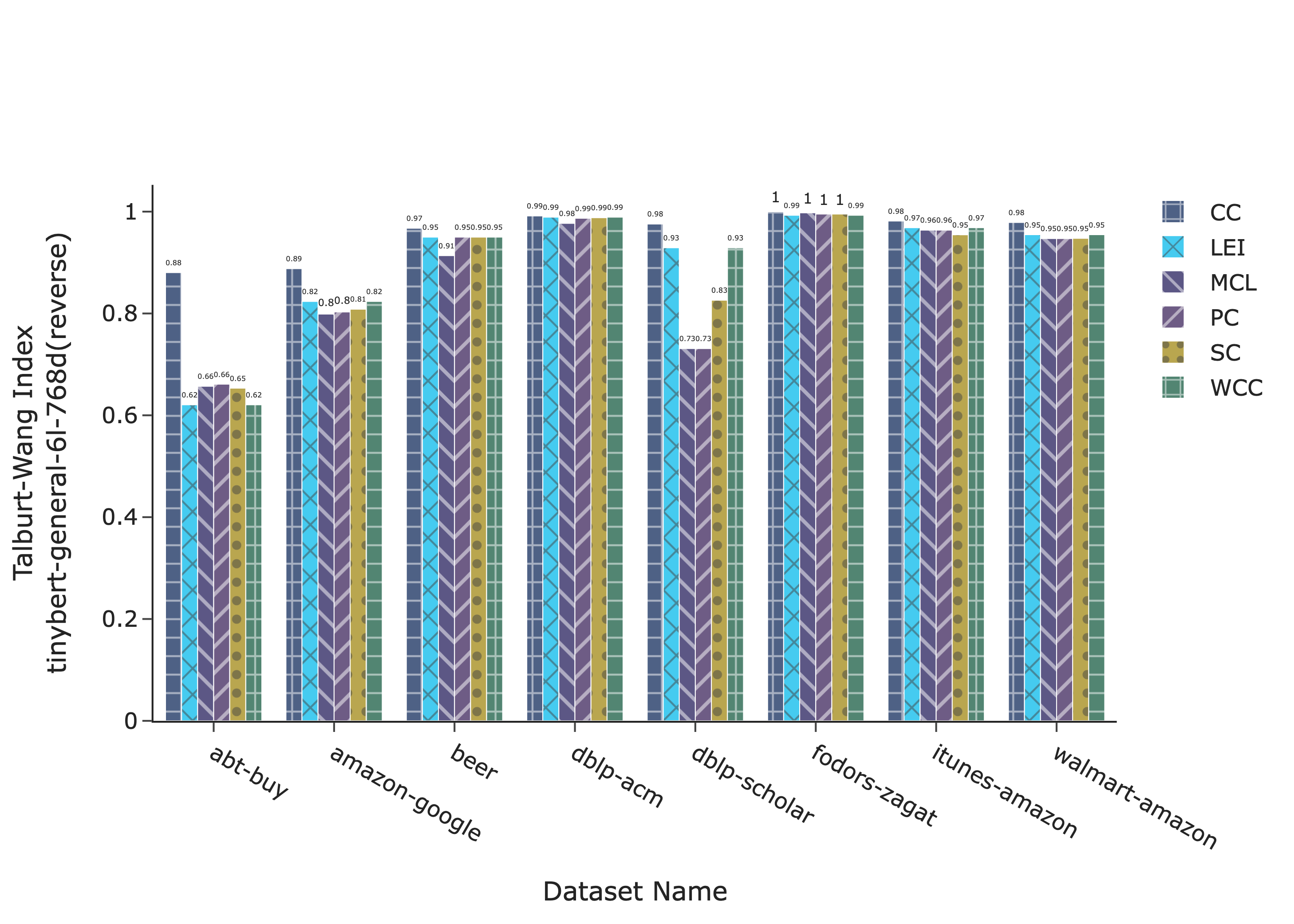}
        \caption{Talburt-Wang Index}
    \end{subfigure}
    \caption{
        Average clustering metrics across all datasets for the TinyBERT model using the
        Reversed input order.
    }\label{fig:tinybert_reverse_avg}
\end{figure*}

Among the directional methods, SC maintains a balanced performance profile owed, at
least in part, to partitioning the reference graph into weak components before
performing spectral clustering. It penalizes the lack of directional signal but avoids
the drastic drops seen in other methods. MCL and PC exhibit a more severe reaction to
the reversed input, heavily penalizing the output quality, particularly on the
\emph{dblp-scholar} dataset. This behavior is quantified in
Figure~\ref{fig:distilbert_reverse_heatmap} and Figure~\ref{fig:tinybert_reverse_avg}.

\subsubsection{Bi-directional Reference Graph}

The bi-directional reference graph is constructed by merging the forward and reverse
graphs. This setup attempts to recover a more complete signal by combining predictions
from both input orders. The results are inconclusive regarding a universal performance
increase, but they reinforce the observation of training method bias. While pairwise
metrics improve in some cases, the overall clustering structure does not consistently
benefit from the merger.

\begin{figure*}[h!]
    \centering
    \begin{subfigure}[b]{0.48\textwidth}
        \includegraphics[width=\linewidth]{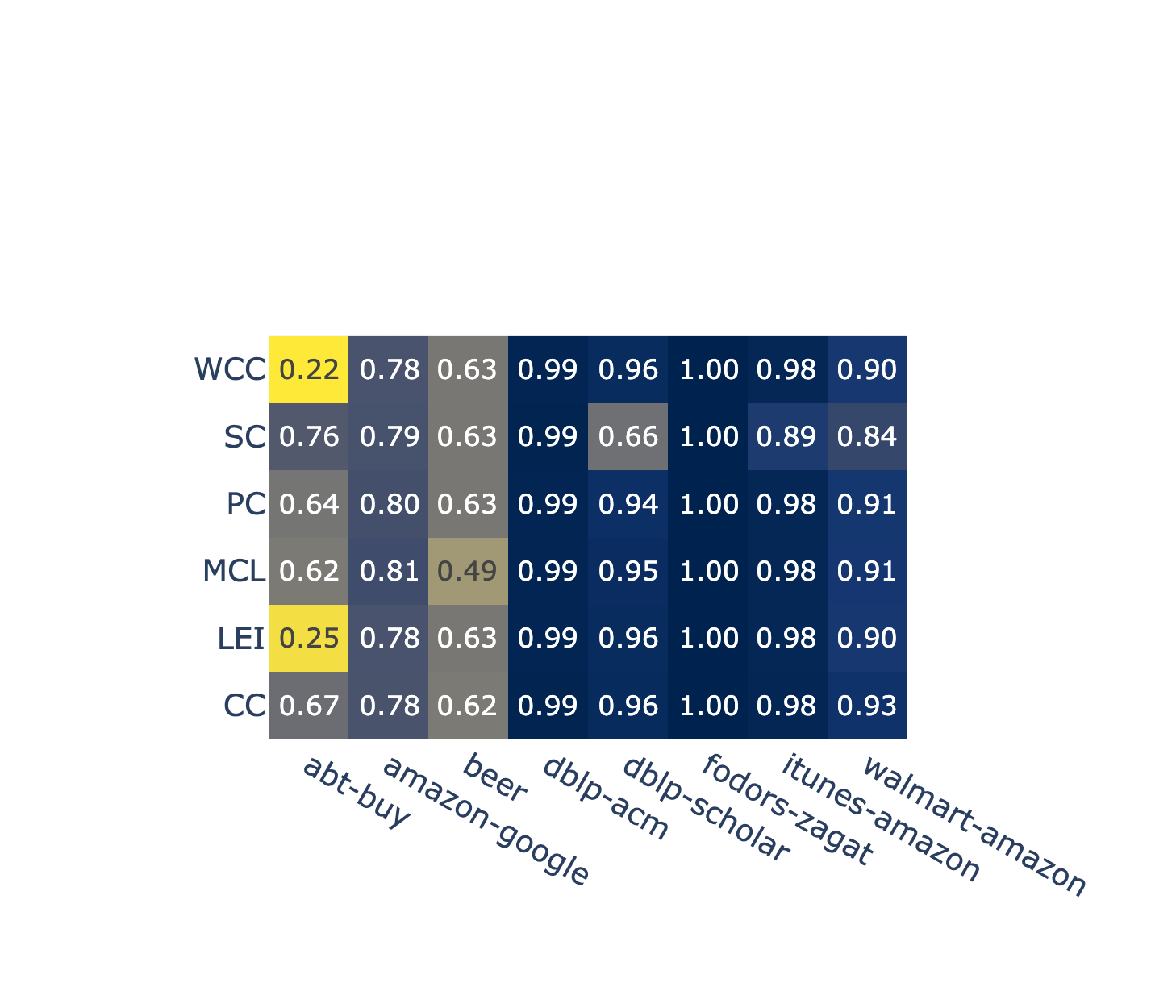}
        \caption{Pair Comparison Measure}
    \end{subfigure}
    \hfill
    \begin{subfigure}[b]{0.48\textwidth}
        \includegraphics[width=\linewidth]{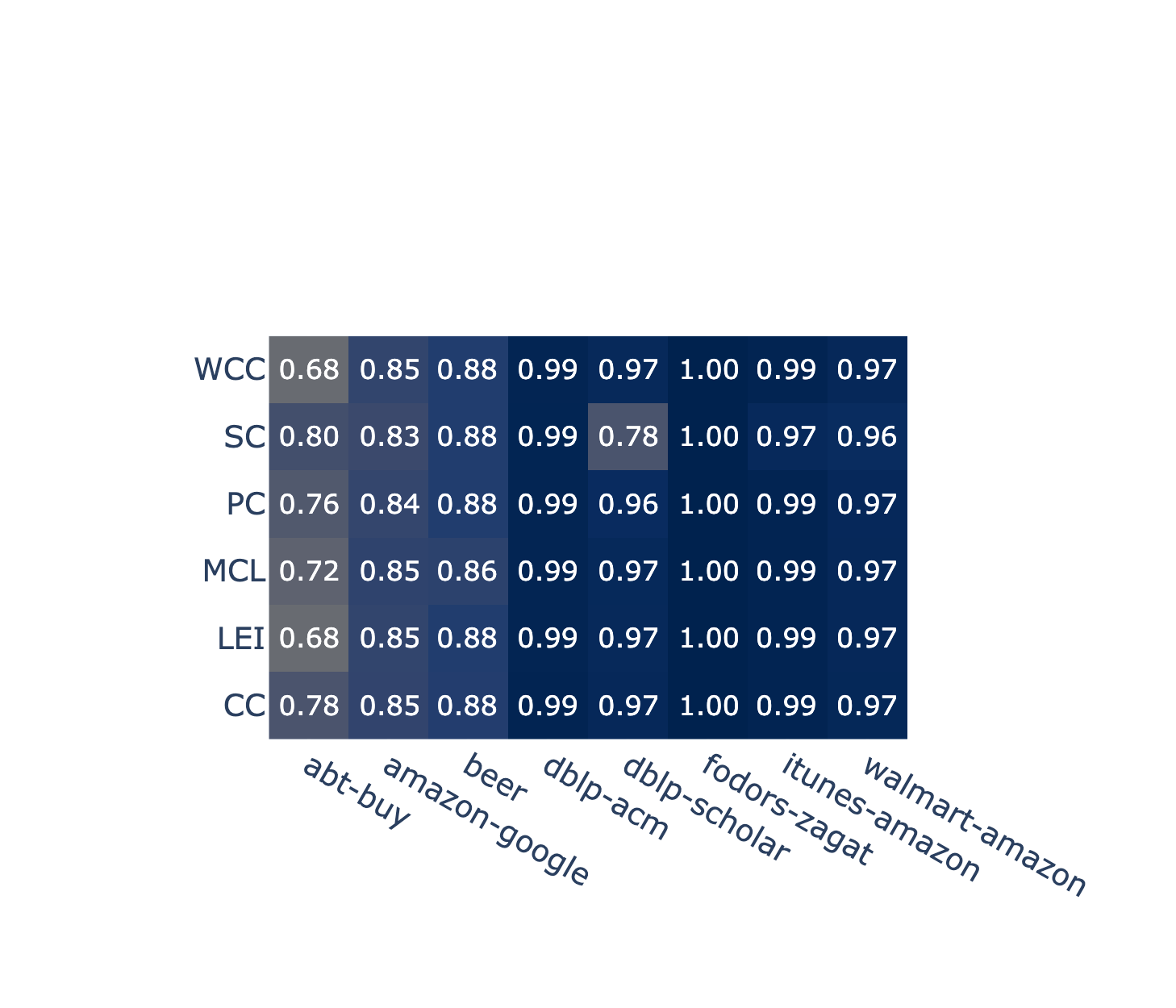}
        \caption{Cluster Comparison Measure}
    \end{subfigure}
    \vfill
    \begin{subfigure}[b]{0.48\textwidth}
        \includegraphics[width=\linewidth]{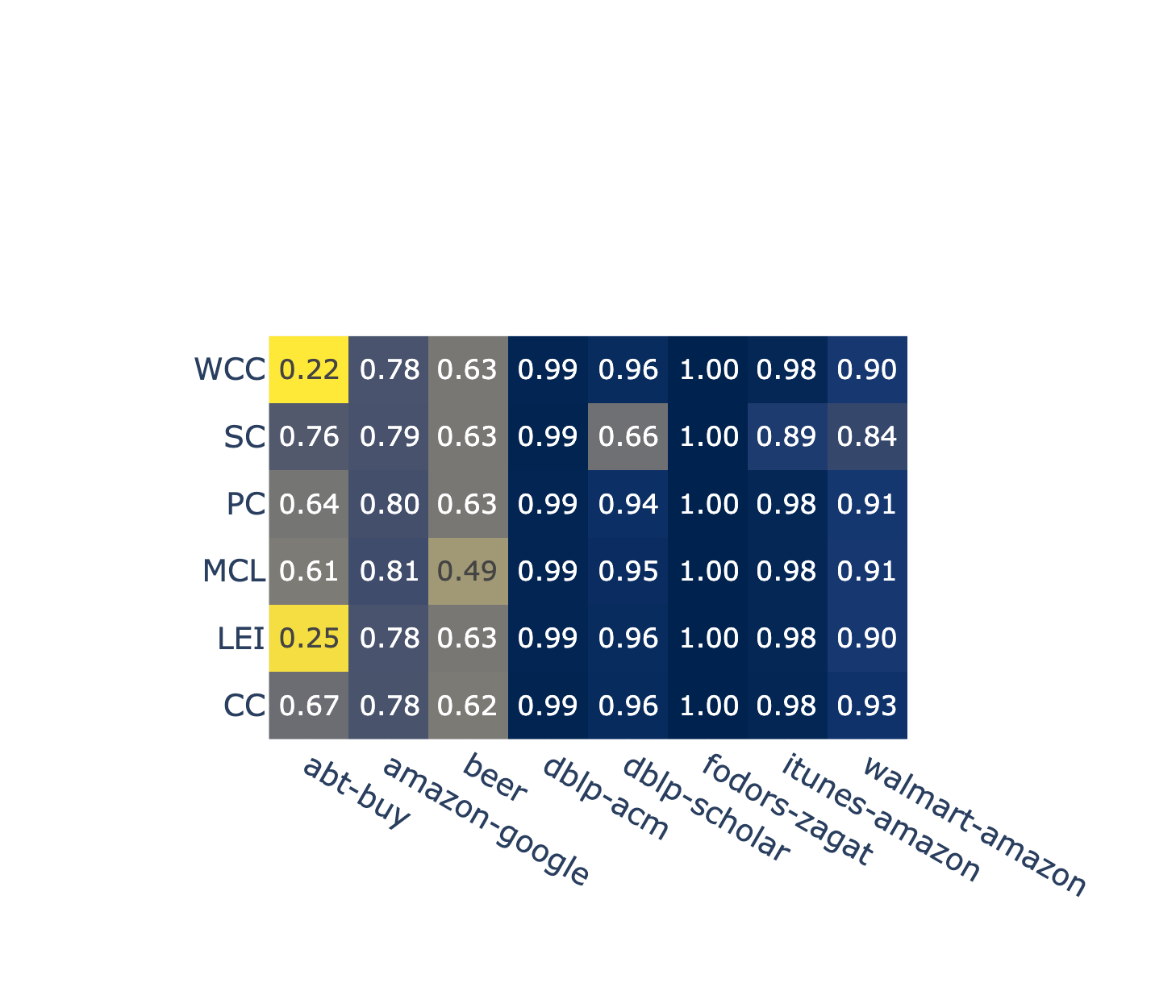}
        \caption{Adjusted Rand Index}
    \end{subfigure}
    \hfill
    \begin{subfigure}[b]{0.48\textwidth}
        \includegraphics[width=\linewidth]{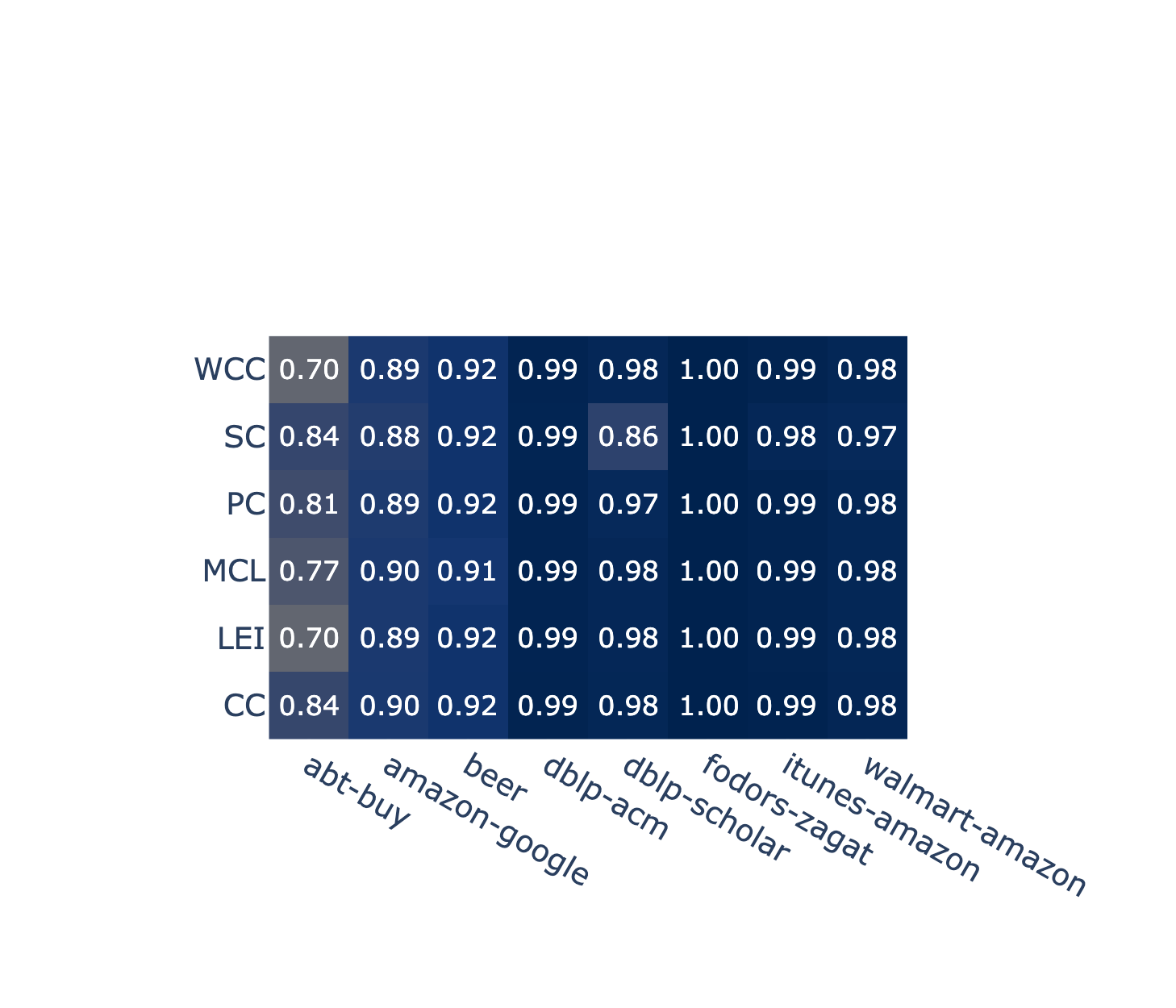}
        \caption{Talburt-Wang Index}
    \end{subfigure}
    \caption{
        Heatmap of clustering metrics for the DeBERTa (v3 base) model using the
        Bi-directional input order. Performance improvements of directional clustering
        algorithms over baselines are visible on textual datasets like \emph{abt-buy},
        while other datasets (\emph{dblp-scholar}, \emph{walmart-amazon}) show
        stagnation or decline.
    }\label{fig:deberta_bidi_heatmap}
\end{figure*}

Specific algorithms show notable performance on difficult textual datasets. SC
demonstrates significant performance gains on \emph{abt-buy} over the CC baseline.
Others (WCC, LEI) show a surprising degradation on the same textual datasets. This
pattern is more pronounced when using the DeBERTa-V3 model for matching
(Figure~\ref{fig:deberta_bidi_heatmap}).

Conversely, performance on \emph{dblp-scholar}, \emph{itunes-amazon}, and
\emph{walmart-amazon} diminishes. This reduction is slight for MCL and PC but drastic
for SC. It remains an open question whether this degradation stems from the matcher
being trained exclusively on a single input order or from latent factors within the
datasets themselves. The disparity suggests that simple bi-directional merging is
insufficient to resolve the asymmetry modeled by the transformers.

\section{Discussion}\label{sec:discussion}

While the experimental results presented in Section~\ref{sec:experiments} highlighted
specific representative scenarios (such as RoBERTa for forward matching and DistilBERT
for reverse scenarios), analyzing the full experimental suite reveals that input-order
asymmetry is not unique to a specific model or dataset.

Across all four transformer architectures tested (RoBERTa, DeBERTa, DistilBERT, and
TinyBERT), we observed consistent performance degradation when comparing the
\emph{Forward} and \emph{Reverse} topologies. While localized variations in F1-scores
between models are attributable to differences in pre-training corpora and parameter
counts, the macroscopic trend remains: the directionality of input serialization
fundamentally alters ER decisions.

Furthermore, the significant performance drop of strict transitivity-enforcing
algorithms (such as SC, PC and MCL) in the \emph{reverse} topology was observed
universally across all model backbones. This confirms that ``bad triplets'' could be
automatically detected and resolved properly using transformer-based matching, thus
reinforcing the necessity for direction-aware clustering approaches, such as the
Leiden, spectral clustering, parent CENTER and Markov clustering methods. These methods
provide higher fidelity when handling the conflicting directional signals generated by
asymmetric matchers.

\subsection*{Data Availability}
To facilitate further analysis, we provide access to the datasets, matcher training
logs and results obtained for all model-dataset combinations via OSF~\cite{olar2026}.

\section{Future Work}\label{sec:e2e-transformers-future-work}

The experimental results presented in this study highlight that current ER pipelines
often suppress, rather than utilize, the inherent asymmetry of transformer-based
matching. To move beyond strict symmetry, future research must focus on three
interconnected areas.

\paragraph{Direction-Aware Training Paradigms.}
The current standard for training neural matchers relies on binary classification
targets (Match vs. Non-Match), effectively forcing the model to symmetrize its
understanding of entity relationships. A promising avenue for future work is to reframe
the matching task as a multi-class problem. Instead of a binary output, the matcher
should be trained to predict specific relationship types: equivalence ($a
\leftrightarrow b$), directional implication or subsumption ($a \to b$ or $b \to a$),
and non-relation. By explicitly modeling directionality, the matcher can distinguish
between strict identity and cases of referential ambiguity, thereby preventing the
propagation of false positives in transitive closures.

\paragraph{Directional Benchmark Creation.}
To support direction-aware training, we must develop datasets that capture the nuances
of ambiguous matching. Existing benchmarks typically enforce global symmetry in their
ground truth, which obscures the ``bad triplet'' problem. Future efforts involve
deriving new benchmarks from existing sources that specifically isolate hierarchical or
ambiguous data patterns. For example, distinguishing between specific entities (e.g.,
``John Doe'', ``Jane Doe'') and ambiguous references (e.g., ``J. Doe''). These datasets
will require custom clustered ground truths that respect the direction of information
flow (e.g., clustering only one of ``John Doe'' or ``Jane Doe'' with ``J. Doe'' based
on directional information), enabling the evaluation of models on their ability to
handle asymmetric entailment.

\paragraph{Exploiting Directed Graphs.}
Finally, the integration of direction-aware matchers will result in intentionally
directed similarity graphs, rather than the incidentally skewed graphs analyzed in this
work. This evolution requires a re-evaluation of the clustering algorithms presented in
this paper. We intend to rerun the directed clustering methods from this study (such as
SC, PC and MCL) on these semantically directed graphs. We hypothesize that by
respecting the directionality of edges—where links represent entailment rather than
just similarity—we can achieve higher clustering precision and potentially uncover
hierarchical structures within entity groups currently hidden by the undirected
reference graphs obtained using symmetric matchers.

\section{Conclusions}\label{sec:e2e-transformers-conclusions}

This extended study generalizes our preliminary findings~\cite{olar2025} to demonstrate
that input-order asymmetry is not an artifact of a specific implementation, but a
fundamental characteristic of transformer-based entity matching. By evaluating a
diverse array of architectures—including DistilBERT, TinyBERT, and DeBERTa—we
established that modern matchers inherently violate the symmetric assumption
traditionally imposed on Entity Resolution (ER).

We addressed this challenge by modeling the ER process as a collective task on directed
reference graphs. Our experiments with direction-aware clustering algorithms, such as
Leiden and Directed Spectral clustering, reveal that preserving asymmetry exposes
significant variations in clustering structure that are otherwise less pronounced when
using directionally unaware community detection methods like Weakly Connected
Components. While enforcing symmetry stabilizes performance on current benchmarks, it
discards the rich directional signal produced by the models. Consequently, we argue
that the path forward lies ``beyond symmetry'': rather than suppressing these
directional cues, future ER systems should adopt direction-aware training paradigms
that explicitly model entailment and ambiguity, turning a theoretical violation into a
practical feature for resolving complex entity relationships.

\begin{credits}
    \subsubsection{\discintname}
    The author has no competing interests to declare that are relevant to the content
    of this article.
\end{credits}

\bibliographystyle{splncs04}
\bibliography{./main}

@inproceedings{olar2025,
  author       = {Andrei Olar},
  booktitle    = {Proceedings of the 17th International Joint Conference on Knowledge Discovery, Knowledge Engineering and Knowledge Management - Volume 1: KDIR},
  doi          = {10.5220/0013672900004000},
  isbn         = {},
  organization = {{INSTICC}},
  pages        = {208-215},
  publisher    = {SciTePress},
  title        = {On the Asymmetrical Nature of Entity Matching Using Pre-Trained Transformers},
  year         = {2025}
}

@article{roberta2019,
  archiveprefix = {arXiv},
  author        = {Yinhan Liu and Myle Ott and Naman Goyal and Jingfei Du and Mandar Joshi and Danqi Chen and Omer Levy and Mike Lewis and Luke Zettlemoyer and Veselin Stoyanov},
  doi           = {10.48550/arXiv.1907.11692},
  eprint        = {1907.11692},
  journal       = {arXiv preprint arXiv:1907.11692},
  primaryclass  = {cs.{CL}},
  title         = {Ro{BERT}a: A Robustly Optimized BERT Pretraining Approach},
  url           = {https://arxiv.org/abs/1907.11692},
  year          = {2019},
  year          = {2019}
}

@article{distilbert2019,
  archiveprefix = {arXiv},
  author        = {Victor Sanh and Lysandre Debut and Julien Chaumond and Thomas Wolf},
  doi           = {10.48550/arXiv.1910.01108},
  eprint        = {1910.01108},
  journal       = {ArXiv},
  primaryclass  = {cs.CL},
  title         = {Distil{BERT}, a distilled version of {BERT}: smaller, faster, cheaper and lighter},
  url           = {https://arxiv.org/abs/1910.01108},
  volume        = {abs/1910.01108},
  year          = {2020}
}

@article{deberta2023,
  archiveprefix = {arXiv},
  author        = {Pengcheng He and Jianfeng Gao and Weizhu Chen},
  doi           = {10.48550/arXiv.2111.09543},
  eprint        = {2111.09543},
  journal       = {ArXiv},
  primaryclass  = {cs.CL},
  title         = {{D}e{BERT}a{V}3: Improving {D}e{BERT}a using {ELECTRA}-Style Pre-Training with Gradient-Disentangled Embedding Sharing},
  year          = {2023}
}

@inproceedings{tinybert2020,
  address   = {Online},
  author    = {Jiao, Xiaoqi  and
               Yin, Yichun  and
               Shang, Lifeng  and
               Jiang, Xin  and
               Chen, Xiao  and
               Li, Linlin  and
               Wang, Fang  and
               Liu, Qun},
  booktitle = {Findings of the Association for Computational Linguistics: EMNLP 2020},
  doi       = {10.18653/v1/2020.findings-emnlp.372},
  editor    = {Cohn, Trevor  and
               He, Yulan  and
               Liu, Yang},
  month     = nov,
  pages     = {4163--4174},
  publisher = {Association for Computational Linguistics},
  title     = {{T}iny{BERT}: Distilling {BERT} for Natural Language Understanding},
  url       = {https://aclanthology.org/2020.findings-emnlp.372/},
  year      = {2020}
}

@article{ditto2020,
  author     = {Li, Yuliang and Li, Jinfeng and Suhara, Yoshihiko and Doan, AnHai and Tan, Wang-Chiew},
  doi        = {10.14778/3421424.3421431},
  issn       = {2150-8097},
  issue_date = {September 2020},
  journal    = {Proc. VLDB Endow.},
  month      = sep,
  number     = {1},
  numpages   = {11},
  pages      = {50-60},
  publisher  = {VLDB Endowment},
  title      = {Deep entity matching with pre-trained language models},
  url        = {https://doi.org/10.14778/3421424.3421431},
  volume     = {14},
  year       = {2020}
}

@article{e2e2020,
  address    = {New York, NY, USA},
  articleno  = {127},
  author     = {Christophides, Vassilis and Efthymiou, Vasilis and Palpanas, Themis and Papadakis, George and Stefanidis, Kostas},
  doi        = {10.1145/3418896},
  issn       = {0360-0300},
  issue_date = {November 2021},
  journal    = {ACM Comput. Surv.},
  month      = dec,
  number     = {6},
  numpages   = {42},
  publisher  = {Association for Computing Machinery},
  title      = {An Overview of End-to-End Entity Resolution for Big Data},
  url        = {https://doi.org/10.1145/3418896},
  volume     = {53},
  year       = {2020}
}

@inproceedings{deepmatcher2018,
  address   = {New York, NY, USA},
  author    = {Mudgal, Sidharth and Li, Han and Rekatsinas, Theodoros and Doan, AnHai and Park, Youngchoon and Krishnan, Ganesh and Deep, Rohit and Arcaute, Esteban and Raghavendra, Vijay},
  booktitle = {Proceedings of the 2018 International Conference on Management of Data},
  doi       = {10.1145/3183713.3196926},
  isbn      = {9781450347037},
  location  = {Houston, TX, USA},
  numpages  = {16},
  pages     = {19-34},
  publisher = {Association for Computing Machinery},
  series    = {SIGMOD '18},
  title     = {Deep Learning for Entity Matching: A Design Space Exploration},
  url       = {https://doi.org/10.1145/3183713.3196926},
  year      = {2018}
}

@inproceedings{huang2006efficient,
  address   = {Berlin, Heidelberg},
  author    = {Huang, Jian
               and Ertekin, Seyda
               and Giles, C. Lee},
  booktitle = {Knowledge Discovery in Databases: PKDD 2006},
  editor    = {F{\"u}rnkranz, Johannes
               and Scheffer, Tobias
               and Spiliopoulou, Myra},
  isbn      = {978-3-540-46048-0},
  pages     = {536--544},
  publisher = {Springer Berlin Heidelberg},
  title     = {Efficient Name Disambiguation for Large-Scale Databases},
  year      = {2006}
}

@article{menestrinaermetrics2010,
  author     = {Menestrina, David and Whang, Steven Euijong and Garcia-Molina, Hector},
  doi        = {10.14778/1920841.1920871},
  issn       = {2150-8097},
  issue_date = {September 2010},
  journal    = {Proc. VLDB Endow.},
  month      = sep,
  number     = {1-2},
  numpages   = {12},
  pages      = {208-219},
  publisher  = {VLDB Endowment},
  title      = {Evaluating entity resolution results},
  url        = {https://doi.org/10.14778/1920841.1920871},
  volume     = {3},
  year       = {2010}
}

@article{bipartitegraphclusters2023,
  author        = {Papadakis, George and Efthymiou, Vasilis and Thanos, Emmanouil and Hassanzadeh, Oktie and Christen, Peter},
  date          = {2023/11/01},
  doi           = {10.1007/s00778-023-00791-3},
  id            = {Papadakis2023},
  isbn          = {0949-877X},
  journal       = {The VLDB Journal},
  number        = {6},
  pages         = {1369--1400},
  title         = {An analysis of one-to-one matching algorithms for entity resolution},
  url           = {https://doi.org/10.1007/s00778-023-00791-3},
  volume        = {32},
  year          = {2023}
}

@inproceedings{badtriplet2008,
  address   = {New York, NY, USA},
  author    = {Ailon, Nir and Charikar, Moses and Newman, Alantha},
  booktitle = {Proceedings of the Thirty-Seventh Annual ACM Symposium on Theory of Computing},
  doi       = {10.1145/1060590.1060692},
  isbn      = {1581139608},
  location  = {Baltimore, MD, USA},
  numpages  = {10},
  pages     = {684-693},
  publisher = {Association for Computing Machinery},
  series    = {STOC '05},
  title     = {Aggregating inconsistent information: ranking and clustering},
  url       = {https://doi.org/10.1145/1060590.1060692},
  year      = {2005}
}

@article{christen2012data,
  author    = {Christen, Peter and Christen, Peter},
  journal   = {Data Matching: Concepts and Techniques for Record Linkage, Entity Resolution, and Duplicate Detection},
  pages     = {229--242},
  publisher = {Springer},
  title     = {Data Matching Systems},
  year      = {2012}
}

@article{fs1969,
  author    = {Ivan P. Fellegi and Alan B. Sunter},
  doi       = {10.1080/01621459.1969.10501049},
  journal   = {Journal of the American Statistical Association},
  number    = {328},
  pages     = {1183--1210},
  publisher = {ASA Website},
  title     = {A Theory for Record Linkage},
  url       = {https://www.tandfonline.com/doi/abs/10.1080/01621459.1969.10501049},
  volume    = {64},
  year      = {1969}
}

@inproceedings{adjrand2001,
  author    = {Ka Yee Yeung and Walter L. Ruzzo},
  journal   = {Bioinformatics},
  number    = {9},
  pages     = {763--774},
  publisher = {Citeseer},
  title     = {Details of the Adjusted Rand index and Clustering algorithms Supplement to the paper ``An empirical study on Principal Component Analysis for clustering gene expression data'' ( to appear in Bioinformatics )},
  url       = {https://api.semanticscholar.org/CorpusID:16021771},
  volume    = {17},
  year      = {2001}
}

@article{hassanzadehclustering2009,
  author     = {Hassanzadeh, Oktie and Chiang, Fei and Lee, Hyun Chul and Miller, Ren\'{e}e J.},
  doi        = {10.14778/1687627.1687771},
  issn       = {2150-8097},
  issue_date = {August 2009},
  journal    = {Proc. VLDB Endow.},
  month      = aug,
  number     = {1},
  numpages   = {12},
  pages      = {1282-1293},
  publisher  = {VLDB Endowment},
  title      = {Framework for evaluating clustering algorithms in duplicate detection},
  url        = {https://doi.org/10.14778/1687627.1687771},
  volume     = {2},
  year       = {2009}
}

@inproceedings{saeediclustering2017,
  address   = {Cham},
  author    = {Saeedi, Alieh
               and Peukert, Eric
               and Rahm, Erhard},
  booktitle = {Advances in Databases and Information Systems},
  editor    = {Kirikova, M{\={a}}r{\={\i}}te
               and N{\o}rv{\aa}g, Kjetil
               and Papadopoulos, George A.},
  isbn      = {978-3-319-66917-5},
  pages     = {278-293},
  publisher = {Springer International Publishing},
  title     = {Comparative Evaluation of Distributed Clustering Schemes for Multi-source Entity Resolution},
  year      = {2017}
}

@article{deeper2018,
  author    = {Ebraheem, Muhammad and Thirumuruganathan, Saravanan and Joty, Shafiq and Ouzzani, Mourad and Tang, Nan},
  doi       = {10.14778/3236187.3236198},
  issn      = {2150-8097},
  journal   = {Proceedings of the VLDB Endowment},
  month     = jul,
  number    = {11},
  pages     = {1454-1467},
  publisher = {Association for Computing Machinery (ACM)},
  title     = {Distributed representations of tuples for entity resolution},
  url       = {http://dx.doi.org/10.14778/3236187.3236198},
  volume    = {11},
  year      = {2018}
}

@misc{markovclusters2025,
  author       = {Guy Allard},
  howpublished = {GitHub repository},
  note         = {Accessed: 2025-03-31},
  title        = {markov\_clustering},
  type         = {software},
  url          = {https://github.com/GuyAllard/markov\_clustering},
  year         = {2025}
}

@article{markovclusters2008,
  author  = {Van Dongen, Stijn},
  doi     = {10.1137/040608635},
  eprint  = {https://doi.org/10.1137/040608635},
  journal = {SIAM Journal on Matrix Analysis and Applications},
  number  = {1},
  pages   = {121-141},
  title   = {Graph Clustering Via a Discrete Uncoupling Process},
  url     = {https://doi.org/10.1137/040608635},
  volume  = {30},
  year    = {2008}
}

@inproceedings{center2000,
  author    = {Taher Haveliwala and Aristides Gionis and Piotr Indyk},
  booktitle = {Third International Workshop on the Web and Databases (WebDB 2000)},
  title     = {Scalable Techniques for Clustering the Web (Extended Abstract)},
  url       = {http://ilpubs.stanford.edu:8090/445/},
  year      = {2000}
}

@article{tarjanunionfind1979,
  author  = {Robert Endre Tarjan},
  doi     = {https://doi.org/10.1016/0022-0000(79)90042-4},
  issn    = {0022-0000},
  journal = {Journal of Computer and System Sciences},
  number  = {2},
  pages   = {110-127},
  title   = {A class of algorithms which require nonlinear time to maintain disjoint sets},
  url     = {https://www.sciencedirect.com/science/article/pii/0022000079900424},
  volume  = {18},
  year    = {1979}
}

@inproceedings{networkx,
  address   = {Pasadena, CA USA},
  author    = {Aric A. Hagberg and Daniel A. Schult and Pieter J. Swart},
  booktitle = {Proceedings of the 7th Python in Science Conference},
  editor    = {Ga\"el Varoquaux and Travis Vaught and Jarrod Millman},
  pages     = {11 - 15},
  title     = {Exploring Network Structure, Dynamics, and Function using NetworkX},
  year      = {2008}
}

@inproceedings{tal2007,
  author = {John R. Talburt and Richard Y. Wang and Kimberly Hess and Emily Kuo},
  doi    = {10.4018/978-1-59904-024-0.ch001},
  title  = {An Algebraic Approach to Data Quality Metrics for Entity Resolution over Large Datasets},
  url    = {https://api.semanticscholar.org/CorpusID:15920003},
  year   = {2007}
}

@article{bhatget2006,
  author    = {Bhattacharya, Indrajit and Getoor, Lise},
  journal   = {Mining graph data},
  publisher = {Wiley Hoboken, NJ},
  title     = {Entity resolution in graphs},
  volume    = {311},
  year      = {2006}
}

@article{bhatget2007,
  address    = {New York, NY, USA},
  author     = {Bhattacharya, Indrajit and Getoor, Lise},
  doi        = {10.1145/1217299.1217304},
  issn       = {1556-4681},
  issue_date = {March 2007},
  journal    = {ACM Trans. Knowl. Discov. Data},
  month      = mar,
  number     = {1},
  numpages   = {36},
  pages      = {5-es},
  publisher  = {Association for Computing Machinery},
  title      = {Collective entity resolution in relational data},
  url        = {https://doi.org/10.1145/1217299.1217304},
  volume     = {1},
  year       = {2007}
}

@inproceedings{bhatget2004,
  author    = {Bhattacharya, Indrajit and Getoor, Lise},
  booktitle = {KDD workshop on link analysis and group detection},
  title     = {Deduplication and group detection using links},
  year      = {2004}
}

@article{pytorch2017,
  author = {Paszke, Adam and Gross, Sam and Chintala, Soumith and Chanan, Gregory and Yang, Edward and DeVito, Zachary and Lin, Zeming and Desmaison, Alban and Antiga, Luca and Lerer, Adam},
  title  = {Automatic differentiation in pytorch},
  year   = {2017}
}

@inproceedings{transformers2020,
  address   = {Online},
  author    = {Thomas Wolf and Lysandre Debut and Victor Sanh and Julien Chaumond and Clement Delangue and Anthony Moi and Pierric Cistac and Tim Rault and Rémi Louf and Morgan Funtowicz and Joe Davison and Sam Shleifer and Patrick von Platen and Clara Ma and Yacine Jernite and Julien Plu and Canwen Xu and Teven Le Scao and Sylvain Gugger and Mariama Drame and Quentin Lhoest and Alexander M. Rush},
  booktitle = {Proceedings of the 2020 Conference on Empirical Methods in Natural Language Processing: System Demonstrations},
  month     = oct,
  pages     = {38-45},
  publisher = {Association for Computational Linguistics},
  title     = {Transformers: State-of-the-Art Natural Language Processing},
  url       = {https://www.aclweb.org/anthology/2020.emnlp-demos.6},
  year      = {2020}
}

@article{wcc1972,
  author   = {R.L. Graham and D.E. Knuth and T.S. Motzkin},
  doi      = {https://doi.org/10.1016/0012-365X(72)90057-X},
  issn     = {0012-365X},
  journal  = {Discrete Mathematics},
  number   = {1},
  pages    = {17-29},
  title    = {Complements and transitive closures},
  url      = {https://www.sciencedirect.com/science/article/pii/0012365X7290057X},
  volume   = {2},
  year     = {1972}
}

@misc{gacl2024,
  archiveprefix = {arXiv},
  author        = {Zihao Li and Dongqi Fu and Hengyu Liu and Jingrui He},
  doi           = {10.48550/arXiv.2412.03008},
  eprint        = {2412.03008},
  primaryclass  = {cs.SI},
  title         = {Provably Extending PageRank-based Local Clustering Algorithm to Weighted Directed Graphs with Self-Loops and to Hypergraphs},
  url           = {https://arxiv.org/abs/2412.03008},
  year          = {2024}
}

@article{winkler2014,
  author   = {Winkler, William E.},
  doi      = {https://doi.org/10.1002/wics.1317},
  eprint   = {https://wires.onlinelibrary.wiley.com/doi/pdf/10.1002/wics.1317},
  journal  = {WIREs Computational Statistics},
  number   = {5},
  pages    = {313-325},
  title    = {Matching and record linkage},
  url      = {https://wires.onlinelibrary.wiley.com/doi/abs/10.1002/wics.1317},
  volume   = {6},
  year     = {2014}
}

@inproceedings{bilenko2003,
  address   = {New York, NY, USA},
  author    = {Bilenko, Mikhail and Mooney, Raymond J.},
  booktitle = {Proceedings of the Ninth ACM SIGKDD International Conference on Knowledge Discovery and Data Mining},
  doi       = {10.1145/956750.956759},
  isbn      = {1581137370},
  location  = {Washington, D.C.},
  numpages  = {10},
  pages     = {39-48},
  publisher = {Association for Computing Machinery},
  series    = {KDD '03},
  title     = {Adaptive duplicate detection using learnable string similarity measures},
  url       = {https://doi.org/10.1145/956750.956759},
  year      = {2003}
}

@article{loster2021knowledge,
  author   = {Loster, Michael and Koumarelas, Ioannis and Naumann, Felix},
  journal  = {Journal of Data and Information Quality},
  month    = {jan},
  number   = 1,
  title    = {Knowledge Transfer for Entity Resolution with Siamese Neural Networks},
  volume   = 13,
  year     = 2021
}

@inproceedings{nseen2019,
  address   = {Berlin, Heidelberg},
  author    = {Fakhraei, Shobeir and Mathew, Joel and Ambite, Jos\'{e} Luis},
  booktitle = {Machine Learning and Knowledge Discovery in Databases: European Conference, ECML PKDD 2019, W\"{u}rzburg, Germany, September 16–20, 2019, Proceedings, Part II},
  doi       = {10.1007/978-3-030-46147-8_40},
  isbn      = {978-3-030-46146-1},
  location  = {W\"{u}rzburg, Germany},
  numpages  = {16},
  pages     = {665–680},
  publisher = {Springer-Verlag},
  title     = {NSEEN: Neural Semantic Embedding for Entity Normalization},
  url       = {https://doi.org/10.1007/978-3-030-46147-8_40},
  year      = {2019}
}

@article{krivosheev2021,
  author  = {Krivosheev, Evgeny and Atzeni, Mattia and Mirylenka, Katsiaryna and Scotton, Paolo and Miksovic, Christoph and Zorin, Anton},
  doi     = {10.1609/aaai.v35i18.18010},
  journal = {Proceedings of the AAAI Conference on Artificial Intelligence},
  month   = {May},
  number  = {18},
  pages   = {16054-16056},
  title   = {Business Entity Matching with Siamese Graph Convolutional Networks},
  url     = {https://ojs.aaai.org/index.php/AAAI/article/view/18010},
  volume  = {35},
  year    = {2021}
}

@article{relater2023,
  address    = {New York, NY, USA},
  articleno  = {12},
  author     = {Kirielle, Nishadi and Christen, Peter and Ranbaduge, Thilina},
  doi        = {10.1145/3533016},
  issn       = {1556-4681},
  issue_date = {January 2023},
  journal    = {ACM Trans. Knowl. Discov. Data},
  month      = feb,
  number     = {1},
  numpages   = {30},
  publisher  = {Association for Computing Machinery},
  title      = {Unsupervised Graph-Based Entity Resolution for Complex Entities},
  url        = {https://doi.org/10.1145/3533016},
  volume     = {17},
  year       = {2023}
}

@article{crosser2026,
  author   = {Yunong Tian and Ning Wang and Anshun Zhou},
  doi      = {10.1016/j.is.2025.102609},
  issn     = {0306-4379},
  journal  = {Information Systems},
  pages    = {102609},
  title    = {CrossER: A robust and adaptable generalized entity resolution framework for diverse and heterogeneous datasets},
  url      = {https://www.sciencedirect.com/science/article/pii/S0306437925000936},
  volume   = {135},
  year     = {2026}
}

@article{grapher2025,
  author   = {Junwei Hu and Michael Bewong and Selasi Kwashie and Yidi Zhang and Vincent Nofong and John Wondoh and Zaiwen Feng},
  doi      = {https://doi.org/10.1016/j.is.2025.102551},
  issn     = {0306-4379},
  journal  = {Information Systems},
  pages    = {102551},
  title    = {When GDD meets GNN: A knowledge-driven neural connection for effective entity resolution in property graphs},
  url      = {https://www.sciencedirect.com/science/article/pii/S0306437925000365},
  volume   = {132},
  year     = {2025}
}

@article{incontext2025,
  address    = {New York, NY, USA},
  articleno  = {252},
  author     = {Fu, Jiajie and Tang, Haitong and Khan, Arijit and Mehrotra, Sharad and Ke, Xiangyu and Gao, Yunjun},
  doi        = {10.1145/3749170},
  issue_date = {September 2025},
  journal    = {Proc. ACM Manag. Data},
  month      = sep,
  number     = {4},
  numpages   = {28},
  publisher  = {Association for Computing Machinery},
  title      = {In-context Clustering-based Entity Resolution with Large Language Models: A Design Space Exploration},
  url        = {https://doi.org/10.1145/3749170},
  volume     = {3},
  year       = {2025}
}

@inproceedings{matchcompareselect2025,
  address   = {Abu Dhabi, UAE},
  author    = {Wang, Tianshu  and
               Chen, Xiaoyang  and
               Lin, Hongyu  and
               Chen, Xuanang  and
               Han, Xianpei  and
               Sun, Le  and
               Wang, Hao  and
               Zeng, Zhenyu},
  booktitle = {Proceedings of the 31st International Conference on Computational Linguistics},
  editor    = {Rambow, Owen  and
               Wanner, Leo  and
               Apidianaki, Marianna  and
               Al-Khalifa, Hend  and
               Eugenio, Barbara Di  and
               Schockaert, Steven},
  month     = jan,
  pages     = {96--109},
  publisher = {Association for Computational Linguistics},
  title     = {Match, Compare, or Select? An Investigation of Large Language Models for Entity Matching},
  url       = {https://aclanthology.org/2025.coling-main.8/},
  year      = {2025}
}

@article{leiden2019,
  author  = {Traag, V. A. and Waltman, L. and van Eck, N. J.},
  doi     = {10.1038/s41598-019-41695-z},
  issn    = {2045-2322},
  journal = {Scientific Reports},
  month   = mar,
  number  = {1},
  pages   = {5233},
  title   = {From {Louvain} to {Leiden}: guaranteeing well-connected communities},
  url     = {https://doi.org/10.1038/s41598-019-41695-z},
  volume  = {9},
  year    = {2019}
}

@article{igraph2006,
  author  = {Csardi, Gabor and Nepusz, Tamas},
  journal = {InterJournal, Complex Systems},
  number  = {5},
  pages   = {1--9},
  title   = {The igraph software package for complex network research},
  url     = {https://igraph.org},
  volume  = {1695},
  year    = {2006}
}

@misc{leidenalg2025,
  author       = {Traag, Vincent A.},
  howpublished = {\url{https://github.com/vtraag/leidenalg}},
  journal      = {GitHub repository},
  note         = {Version 0.12.0},
  publisher    = {GitHub},
  title        = {leidenalg: Implementation of the {Leiden} algorithm for various quality functions in {C}++},
  year         = {2025}
}

@article{chung2005laplacians,
  author    = {Chung, Fan},
  doi       = {10.1007/s00026-005-0237-z},
  journal   = {Annals of Combinatorics},
  number    = {1},
  pages     = {1--19},
  publisher = {Springer},
  title     = {Laplacians of directed graphs},
  volume    = {9},
  year      = {2005}
}

@article{vonluxburg2007tutorial,
  author    = {Von Luxburg, Ulrike},
  doi       = {10.1007/s11222-007-9033-z},
  journal   = {Statistics and Computing},
  number    = {4},
  pages     = {395--416},
  publisher = {Springer},
  title     = {A tutorial on spectral clustering},
  volume    = {17},
  year      = {2007}
}

@inproceedings{ng2002spectral,
  author    = {Ng, Andrew Y and Jordan, Michael I and Weiss, Yair},
  booktitle = {Advances in Neural Information Processing Systems},
  pages     = {849--856},
  publisher = {MIT Press},
  title     = {On spectral clustering: Analysis and an algorithm},
  volume    = {14},
  year      = {2002}
}

@techreport{pagerank1999,
  author      = {Lawrence Page and Sergey Brin and Rajeev Motwani and Terry Winograd},
  institution = {Stanford InfoLab},
  month       = {November},
  note        = {Previous number = SIDL-WP-1999-0120},
  number      = {1999-66},
  publisher   = {Stanford InfoLab},
  title       = {The PageRank Citation Ranking: Bringing Order to the Web.},
  type        = {Technical Report},
  url         = {http://ilpubs.stanford.edu:8090/422/},
  year        = {1999}
}

@article{malliaros2013clustering,
  author    = {Malliaros, Fragkiskos D and Vazirgiannis, Michalis},
  doi       = {10.1016/j.physrep.2013.08.002},
  journal   = {Physics Reports},
  number    = {4},
  pages     = {95--142},
  publisher = {Elsevier},
  title     = {Clustering and community detection in directed networks: A survey},
  volume    = {533},
  year      = {2013}
}

@article{leicht2008community,
  author    = {Leicht, Elizabeth A. and Newman, Mark E. J.},
  doi       = {10.1103/PhysRevLett.100.118703},
  journal   = {Physical Review Letters},
  number    = {11},
  pages     = {118703},
  publisher = {American Physical Society},
  title     = {Community structure in directed networks},
  volume    = {100},
  year      = {2008}
}

@misc{olar2026,
  author    = {Olar, Andrei},
  doi       = {10.17605/OSF.IO/MQ2N7},
  month     = {Feb},
  publisher = {OSF},
  title     = {Beyond Symmetry: Asymmetric Matching with Transformers for Collective Entity Resolution - Experiment Data},
  url       = {https://osf.io/mq2n7/overview?view_only=179bf44ae66c4b26be1e77916a7146b3},
  year      = {2026}
}

\end{document}